\documentclass[a4paper, 11pt, oneside]{article}
\pdfoutput=1
\usepackage{jheppub}
\usepackage{graphicx,wrapfig,float,slashed,cancel}
\usepackage{amsmath,amssymb}
\allowdisplaybreaks	
\usepackage{bbm}
\usepackage{nicefrac}
\usepackage{sidecap}	

\newcommand{\nc}{\newcommand}
\nc{\non}{\nonumber}
\nc{\hc}{\hbox {h.c.}}
\nc{\noi}{\noindent}
\nc{\barx}{\bar{x}}
\nc{\pbarn}{\;\hbox {pb}}
\nc{\fbarn}{\;\hbox {fb}}

\nc{\hsp}{\hspace{0.4cm}}
\nc{\lsp}{\hspace{0.8cm}}
\nc{\Lsp}{\hspace{1.6cm}}
\nc{\what}{\widehat}
\nc{\LLsp}{\lsp\lsp}
\nc{\lra}{\longrightarrow}
\nc{\p}{\prime}
\nc{\sgn}{\text{sgn}}
\nc{\ph}{\varphi}
\nc{\op}{{\cal O}}
\nc{\one}[1]{\mathbbm{1}_{#1}}
\nc{\cL}{\mathcal{L}}
\nc{\mo}{\texttt{micrOMEGAs }}
\nc{\vmol}{v_{\text{M\o l}}}
\nc{\eff}{\text{eff}}
\nc{\sm}{\text{\rm SM}}
\nc{\dm}{\text{DM}}
\nc{\mchi}{m_\chi}
\nc{\mphi}{m_\phi}
\nc{\half}{\nicefrac12}

\newcommand{\kom}[1]{}

\renewcommand{\O}{\mathcal{O}}
\newcommand{\M}{\mathcal{M}}

\let\oldbf\bf\renewcommand{\bf}{\boldmath\oldbf}
\newcommand{\gev}{\text{\,GeV}}

\newcommand{\Z}{\mathbb{Z}}

\newcommand{\R}{\mathcal{R}}
\newcommand{\X}{\mathcal{X}}
\newcommand{\lag}{\mathcal{L}}

\newcommand{\pd}{\partial}
\newcommand{\pds}{\slashed\partial}
\newcommand{\conj}{{\raisebox{-2pt}{$\scriptstyle{c}$}}}
\newcommand{\chic}{\chi^\conj}

\newcommand{\yx}{y_{_X}}
\newcommand{\gx}{g_{_X}}
\newcommand{\vs}{v_S}
\newcommand{\at}[1]{\Big{|}_{\raisebox{-3pt}{$\scriptstyle{#1}$}}}

\newcommand{\mdm}{m_\dm}
\newcommand{\lsim}{\mathrel{\raise.3ex\hbox{$<$\kern-.75em\lower1ex\hbox{$\sim$}}}}
\newcommand{\br}{\text{BR}}

\def\bal#1\eal{\begin{align} #1 \end{align}}
\nc{\bald}{\begin{aligned}}  \nc{\eald}{\end{aligned}}
\nc{\beq}{\begin{equation}}  \nc{\eeq}{\end{equation}}
\nc{\bea}{\begin{eqnarray}}  \nc{\eea}{\end{eqnarray}}
\def\lsim{\mathrel{\raise.3ex\hbox{$<$\kern-.75em\lower1ex\hbox{$\sim$}}}}
\def\gsim{\mathrel{\raise.3ex\hbox{$>$\kern-.75em\lower1ex\hbox{$\sim$}}}}
\renewcommand{\Re}{\mbox{Re\thinspace}}
\renewcommand{\Im}{\mbox{Im\thinspace}}

\def\gev{\;\text{GeV}}

\def\z2{\Z_2}
\def\lamh{\lambda_H}
\def\lams{\lambda_S}

\def\vs{v_S}
\def\va{v_A}
\def\mx{m_{_X}}

\def\uone{U(1)_X}
\def\inv#1{\frac1{#1}}
\newcommand{\sdm}{\text{pGDM}}
\newcommand{\fdm}{\text{FDM}}
\newcommand{\vdm}{\text{VDM}}
\newcommand{\Mdnd}[4]{\left[\!\!\begin{array}{cc}#1&#2\\#3&#4\end{array}\!\!\right]}

\title{Dark-matter-spin effects at future {\boldmath $e^+e^-$} colliders}
\author{Bohdan Grzadkowski,} 
\author{Michal Iglicki,} 
\author{Krzysztof Mekala} 
\author{and Aleksander Filip Zarnecki}
\affiliation{Faculty of Physics, University of Warsaw, Pasteura 5, 02-093 Warsaw, Poland}
\emailAdd{bohdan.grzadkowski@fuw.edu.pl}
\emailAdd{michal.iglicki@fuw.edu.pl}
\emailAdd{k.mekala@student.uw.edu.pl}
\emailAdd{filip.zarnecki@fuw.edu.pl}

\abstract{
We discuss the possibility to detect spin 0, 1 and $\half$ dark matter (DM) at future $e^+e^-$ colliders. The models considered here are simple, consistent and renormalizable field theories that provide correct DM abundance and satisfy direct detection, indirect detection and collider constraints. The intention of this paper was to verify to what extent it might be possible to disentangle models of different DM spins by the measurement of the cross section for $e^+e^-\to Z+\cdots$ at future $e^+e^-$ colliders. We specialize to the case of the ILC operating at $\sqrt{s}=250\gev$, however our results apply as well for the FCC-ee and the CEPC colliders. For each model the cross section maximized with respect to parameters was calculated and compared to the expected 95\% CL cross-section limits estimated for the ILC. It turned out that near the $2 \mdm \simeq m_{1,2}$ resonances, where $m_1$ and $m_2$ are the SM Higgs boson and a non-standard Higgs boson masses, respectively, there exist substantial regions where the models are testable. A special attention has been payed to calculation of the cross section in the region where $m_1\simeq m_2$.   
}


\begin{document}

\maketitle
\flushbottom

\section{Introduction} 
\label{intro}

In spite of the Higgs-boson discovery at CERN's Large Hadron Collider (LHC) by the ATLAS~\cite{ATLAS:2012ae} and CMS~\cite{Chatrchyan:2012tx} collaborations, the underlying theory of fundamental interactions is still missing since the Standard Model (SM) does not provide a candidate for dark matter (DM), while its existence has been confirmed by many independent experiments (see e.g. \cite{Zwicky:1933gu, Corbelli:1999af, Dar:1992hc, Clowe:2003tk, Bertone:2004pz, Bartelmann:1999yn, Sofue:2000jx, Ade:2015xua, Hinshaw:2012aka}). 
In this project we are going to discuss minimal extensions of the SM that describe dark matter of various spins (0, 1, \half) in a framework of a consistent, renormalizable quantum field theory. Even if the ultimate theory of DM will prove to be non-minimal, it is reasonable to expect that the minimal models discussed here will capture its major low-energy properties. Our intention is to verify to what extent future $e^+e^-$ colliders operating near $\sqrt{s}=250\gev$: the Future Circular Collider (FCC-ee)~\cite{Abada:2019zxq,Blondel:2019qlh,Blondel:2019yqr}, the Circular Electron Positron Collider (CEPC)~\cite{CEPCStudyGroup:2018ghi} and the International Linear Collider (ILC)~\cite{Fujii:2017ekh,Fujii:2017vwa,Bambade:2019fyw}, could be useful for detecting DM in the process of mono-$Z$ production, $e^+e^-\to Z + \cdots$. 
Our strategy is to impose existing constraints on simple models of pseudoscalar (pGDM), vector (VDM) and fermion (FDM) dark matter and determine regions of parameters in which the DM-production cross section at $e^+e^-$ colliders is maximal. Then we compare the maximized predictions with the expected 95\% CL cross-section limits at the ILC, assuming that it will provide a satisfactory estimate for the other colliders as well. That way we are trying to verify whether the future electron-positron colliders operating in the vicinity of $\sqrt{s}=250\gev$ could be used to  test theories of DM.

DM production at future $e^+e^-$ colliders has already been discussed in the literature, see \cite{Kamon:2017yfx,Ko:2016xwd}. However, our approach has another motivation, also the models adopted here are not the same. The goal of this project is different as well.
 
The paper is organized as follows: after the introduction in section~\ref{intro}, in the subsequent sections~\ref{pGDM}, \ref{VDM} and \ref{FDM} we describe the pseudo-Goldstone, vector and fermion dark matter models, respectively. Section~\ref{astro_con} is devoted to the constraints on dark matter scenarios, that are adopted in the paper. In section~\ref{Production} we calculate the cross section for the $e^+e^-\to Z + \cdots$ process and discuss subtleties of the mass degenerate case, $m_1\simeq m_2$. The next section, section~\ref{coll_exp}, is to review the expected sensitivity to this process at the ILC. Section~\ref{num_res} contains our numerical results with determination of regions in the parameter space that could be tested at the FCC-ee, CEPC and ILC.
In the final section, section~\ref{sec:summary}, we summarize our findings. In appendices we collect results concerning the Higgs boson decay widths and 2-point 1-loop scalar Green's functions.

\section{Pseudo-Goldstone dark matter}
\label{pGDM}

In spite of the fact that the minimal model of scalar (spin zero) DM~\cite{Silveira:1985rk,McDonald:1993ex} assumes merely an addition of a real scalar field odd under a $\z2$ symmetry, here we are going to consider a model (pGDM) that requires an extension by a complex scalar filed $S$. The model is in some sense very similar to vector and fermion dark matter models that will be discussed here as well, so it is worth to compare all of them.
In order to stabilize a component of $S$ we require an invariance under DM charge conjugation $C:\; S\rightarrow S^*$, which guarantees stability of the imaginary part of $S$, $A\equiv \Im S/\sqrt{2}$. The real part, $\phi_S \equiv \Re S/\sqrt{2}$, is going to develop a real vacuum expectation value (vev) $\langle \phi_S \rangle = \langle S \rangle = v_S/\sqrt{2}$.\footnote{This is a choice that fixes the freedom (phase rotation of the complex scalar) of choosing a weak basis that could be adopted to formulate the model. The model is defined by symmetries imposed in this particular basis in which the scalar vacuum expectation value is real.} Therefore, $\phi_S$ will mix with the neutral component of the SM Higgs doublet $H$, in exactly the same manner as it happens for the VDM or the FDM. In order to simplify the potential we impose in addition a $\z2$ symmetry $S \rightarrow -S$, which eliminates odd powers of $S$.
Eventually, the scalar potential reads:
\begin{equation}
	V=-\mu_H^2 |H|^2+\lambda_H |H|^4 -\mu_S^2|S|^2 + \lambda_S|S|^4+\kappa|S|^2|H|^2 + \mu^2 (S^2+S^{*\, 2})
	\label{pot_sdm}
\end{equation}
with $\mu^2$ real, as implied by the $C$ symmetry. 
Note that the $\mu^2$ term breaks the $U(1)$ explicitly, so the pseudo-Goldstone boson $A$ is massive. In the limit of exact symmetry, $A$ would be just a genuine, massless Goldstone boson. Since the symmetry-breaking operator $\mu^2 (S^2+S^{*\, 2})$ is of dimension less that 4, its presence does not jeopardize renormalizability even if non-invariant higher dimension operators were not introduced, see for instance \cite{Pokorski:1987ed}. 
Note that dimension 3 terms are disallowed by the $\z2$ and gauge symmetries. In other words, we can limit ourself to dimension 2 $U(1)$-breaking terms preserving the renormalizability of the model. The freedom to introduce solely the soft breaking operators offers a very efficient and economical way to generate mass for the pseudo-scalar $A$ without the necessity to introduce dimension 4 terms like $S^4$ or $|S|^2S^2$, and keeping the renormalizability of the model.
It is also worth noticing that the $\z2$ symmetry $S \rightarrow -S$ is broken spontaneously by $v_S$ and, therefore, $\phi_S$, the real part of $S$, is not stable, making $A$ the only DM candidate.

The scalar fields can be expanded around the corresponding generic vevs, $v$ for $H$ and $\vs$ for $S$, as follows:
	\bal
	S&=\frac{1}{\sqrt{2}}(\vs + i \va + \phi_S + iA)\;,&
	H^0&= \frac{1}{\sqrt{2}}(v + \phi_H+ i\sigma_H)&
	&\text{where }H=\binom{H^+}{H^0}\;.
	\eal
The global minimum of the potential with corresponding value of the potential and the scalar mass-squared matrix read:
	\bea
	v^2&=&\frac{4 \lambda_S \mu^2_H - 2\kappa (\mu^2_S-2 \mu^2)}{4\lambda_H\lambda_S-\kappa^2}\;,\hsp
	v_S^2=\frac{4 \lambda_H (\mu^2_S-2\mu^2) - 2\kappa \mu^2_H }{4\lambda_H\lambda_S-\kappa^2}\;,\hsp \va^2=0 \hsp\label{vac1}\\
	V_\text{min}&=&\frac{-1}{4\lambda_H\lambda_S-\kappa^2}\left\{\lamh(\mu^2_S-2\mu^2)^2+\mu_H^2\left[\lams\mu_H^2-\kappa (\mu^2_S-2\mu^2)
	\right]\right\}\;,\label{V1}\\
	\mathcal{M}^2 &=& \left( 
	\begin{array}{ccc}
	2 \lambda_H v^2  & \kappa v v_S & 0 \\ 
	 \kappa v v_S   & 2 \lambda_S v^2_S & 0 \\
	0 & 0 & - 4 \mu^2
	\end{array} 
	\right) \label{massv1}
	\eea
in the basis $(\phi_H,\phi_S,A)$. Note that the third spin-zero state $A$ does not mix with the former ones. 

Conditions necessary to guarantee the asymptotic positivity of the potential and the global minimum at $(v_H/\sqrt{2}, \vs/\sqrt{2})$ with non-zero vevs will be discussed in section~\ref{theo_con}.

It is worth to notice that in the vector DM model considered in the following section, $A$ becomes a genuine Goldstone boson ($\mu^2=0$) and disappears as a longitudinal component of the massive DM vector $X$.

There are two mass eigenstates, $h_1$ and $h_2$, in this model. The mass matrix (\ref{massv1}) can be diagonalized by the orthogonal rotation matrix $\R^{-1}$ acting on the space spanned by the two CP-even scalars $\phi_H$ and $\phi_S$:
	\begin{eqnarray}\label{mixing_s}
	\left(\begin{array}{c}
	h_1 \\ h_2
	\end{array}\right)=
	\R^{-1}
	\left( \begin{array}{c}
	\phi_H \\ \phi_S
	\end{array}\right)
	=
	\left(\begin{array}{cc}
	\cos\alpha & \sin\alpha \\
	-\sin\alpha & \cos\alpha
	\end{array} \right)
	\left( \begin{array}{c}
	\phi_H \\ \phi_S
	\end{array}\right) \,,
	\label{mix}
	\end{eqnarray}
with
	\beq
	\tan 2\alpha=\frac{2 {\cal M}_{12}^2}{{\cal M}_{11}^2-{\cal M}_{22}^2} \,.
	\label{mix_alp}
	\eeq
We assume hereafter that $h_1$ is the $125.09\gev$ boson observed at the LHC. Moreover, since $\sin\alpha$ appears in calculations of sections~\ref{astro_con} and \ref{Production} in the second power, we will assume without losing generality that the sign of $\kappa$ is chosen in such a way that $\sin\alpha>0$.

We choose as independent parameters of the model the set: $v_S$, $\sin\alpha$, $m_2$ and $\mdm=m_A$. Together with $v=246.22\gev$ and $m_1=125.09\gev$ this set is sufficient to determine all the 6 parameters of the potential; relevant relations will be presented in section~\ref{theo_con}. As it will be seen later, scalar potentials in other theories discussed in this work could be also parametrized in terms of the same parameters, allowing for meaningful comparison between the models.\footnote{Here, the DM mass $m_A$ is also a parameter of the potential. In the remaining models discussed in this paper, DM masses will be independent parameters.}

\begin{figure}[!ht]
\begin{center}\includegraphics[height=0.2\textwidth]{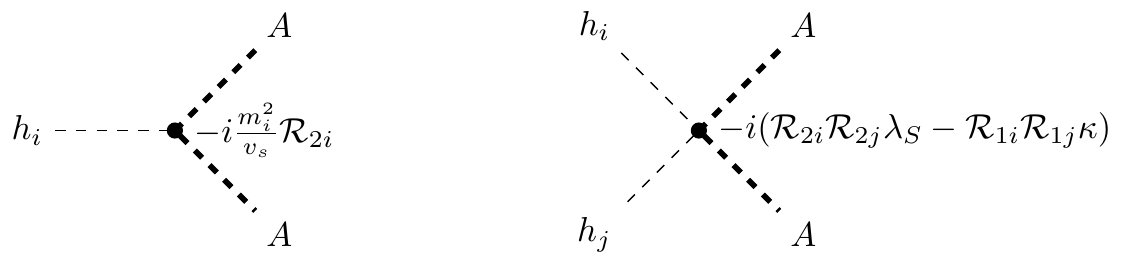}\end{center}
\caption{Vertices relevant for the pGDM model.}
\label{SDM_ver}
\end{figure}

Vertices relevant for the calculation of annihilation cross section in the pGDM model have been collected in figure~\ref{SDM_ver}.

Similar models have been considered in a more general context including a possibility of fast first-order phase transition in \cite{Gonderinger:2012rd, Barger:2010yn, Barger:2008jx}. 
However, those models have different phenomenology, as the pGDM model possesses the unique and attractive feature of natural suppression of DM scattering against nuclei. 
This property of the pGDM is a consequence of the  particular way of soft breaking of the $\uone$ by the terms that are quadratic in $S$, see \cite{Azevedo:2018oxv}.
This aspect will be particularly relevant in section~\ref{DD_dd_vdm}. 

\section{Vector dark matter}
\label{VDM}

The next model that we want to compare with the pGDM is the popular vector DM (VDM) model~\cite{Hambye:2008bq,Lebedev:2011iq,Farzan:2012hh,Baek:2012se,Baek:2014jga,Duch:2015jta} 
that is an extension of the SM by an additional $\uone$ gauge symmetry and a complex scalar field 
$S$, whose vev generates a mass for the corresponding gauge field. The quantum numbers of the scalar field are 
\beq
S  =   (0,{\mathbf{1}},{\mathbf{1}},1) \ \  \text{under}  \ \  U(1)_Y\times SU(2)_L \times SU(3)_c \times \uone.
\eeq
None of the SM fields are charged under the extra gauge group. In order to ensure stability of the new vector boson a $\Z_2$ symmetry is assumed to forbid $U(1)$-kinetic mixing between $\uone$ and  $U(1)_Y$. The extra gauge boson $X$ and the scalar field $S$ transform under the $\Z_2$ as follows
\beq
	X \rightarrow -X\;,\lsp
	S \rightarrow S^*\;.
\eeq
All other fields are neutral under the $\Z_2$. 

The vector bosons' masses are given by:
	\bal
	m_W &= \inv2 g v\;,&
	m_Z &= \inv2 \sqrt{g^2+g'^2}\;v&
	&\text{and}& \mx &= \gx \vs,
	\eal
where $g$ and $g'$ are the $SU(2)$ and $U(1)$ gauge couplings, while, as in the previous model, $v$ and $\vs$ are the vevs of $H$ and $S$, respectively: $(\langle H \rangle,\langle S \rangle)=\frac{1}{\sqrt{2}}(v,\vs)$.\footnote{$\langle H \rangle$ and $\langle S \rangle$ could be chosen to be real and non-negative without losing generality.}
The scalar potential for this model is given by
\beq
V= -\mu^2_H|H|^2 +\lambda_H |H|^4 -\mu^2_S|S|^2 +\lambda_S |S|^4 +\kappa |S|^2|H|^2 .
\label{pot_vdm}
\eeq
It is easy to find solutions of the potential minimization conditions for the scalar fields:
	\bal
	v^2&=\frac{4 \lambda_S \mu^2_H - 2\kappa \mu^2_S }{4\lambda_H\lambda_S-\kappa^2}\;,&
	\vs^2&=\frac{4 \lambda_H \mu^2_S - 2\kappa \mu^2_H }{4\lambda_H\lambda_S-\kappa^2}\;.
	\label{min_con}
	\eal

Both scalar fields can be expanded around the corresponding vevs as follows
	\bal
	S &= \frac{1}{\sqrt{2}}(\vs+ \phi_S +i\sigma_S)\;,&
	H^0 &= \frac{1}{\sqrt{2}}(v + \phi_H+ i\sigma_H)&
	&\text{where}&
	H&=\binom{H^+}{H^0}\;.
	\eal
The mass-squared matrix $\mathcal{M}^2$ for the fluctuations $ \left(\phi_H, \phi_S\right)$ is identical as the $2 \times 2$ block of the mass matrix for the pGDM model (\ref{massv1}), so that the diagonalization \ref{mix} and relation (\ref{mix_alp}) remain applicable. 

Conditions for existence of non-zero vevs, globality of the minimum and asymptotic positivity of the potential will be discussed in section~\ref{theo_con}. The input parameters adopted here are: $v_S$, $\sin\alpha$, $m_2$ and $\mdm=\mx$.

\begin{figure}[!ht]
	\centering
	\includegraphics[height=0.2\textwidth]{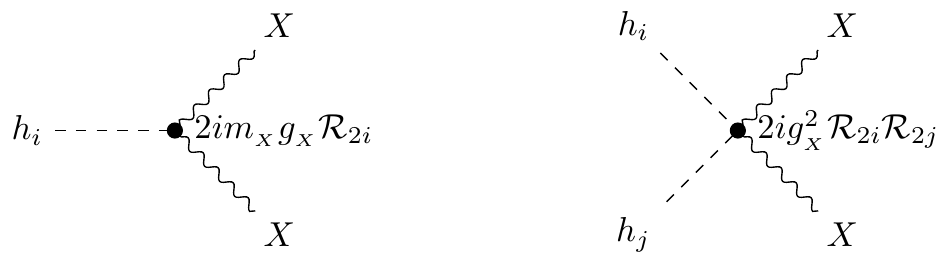}
	\caption{The vertices relevant for the VDM model.}
	\label{fig:VDM_ver}
	\end{figure}

Vertices relevant for the calculation of annihilation cross section in the VDM model have been collected in figure~\ref{fig:VDM_ver}.
It is interesting to notice similarity between the VDM and the pGDM. In the latter one the $\uone$ (that is a gauge symmetry of the VDM) is explicitly (but softly) broken. The corresponding pseudo-Goldstone boson $A$ in the pGDM model remains in the spectrum of scalars, while in the VDM this degree of freedom disappears as a longitudinal component of the massive vector $X$.
\section{Fermion dark matter}
\label{FDM}

In the case of minimal fermion DM, the gauge group remains the standard one, i.e. $U(1)_Y \times SU(2)_L \times SU(3)_c$. This model can be treated as a special case of the singlet-singlet model discussed in \cite{Freitas:2015hsa}.
The DM candidate $\chi$ (left-handed Dirac fermion) is introduced together with a real scalar $S$ that is necessary to mediate DM interaction with the SM.

The extra states are charged under $\Z_4$: $S\to -S$ while $\chi\to i\chi$. The resulting symmetric Lagrangian reads:
\bea
		\lag &=&\lag_\sm+i\bar\chi\pds\chi+\frac{1}{2}\pd^\mu S\,\pd_\mu S-\frac{\yx}{2}(\bar\chic\chi+\bar\chi\chic)S-V(H,S)\;,\label{lag_fdm}\\
		V(H,S)&=&-\mu_H^2|H|^2+\lambda_H|H|^4-\frac{\mu_S^2}{2}S^2+\frac{\lambda_S}{4}S^4+\frac{\kappa}{2}|H|^2S^2\;,\label{pot_fdm}
\eea
where $\chi^\conj\equiv-i\gamma_2\chi^*$. Note that the above potential is the same as in the VDM case (see~\ref{pot_vdm}), up to normalization of the singlet mass and its couplings. The positivity conditions of the potential remain, of course, the same for this model as for the previous two since all the potentials have the same asymptotic behaviour.

We parametrize fluctuations of scalar fields as follows:
	\bal
	S &= \vs+ \phi_S\;,&
	H^0 &= \frac{1}{\sqrt{2}}(v + \phi_H+ i\sigma_H)&
	&\text{where}&
	H&=\binom{H^+}{H^0}\;,
	\eal
with $v$ and $\vs$ being the vevs of the neutral component of the doublet $H$ and the singlet $S$, respectively, determined by (\ref{min_con}). 

After SSB, relevant parts of the Lagrangian take the following form:
	\beq
	i\bar\chi\pds\chi+\frac{1}{2}\pd^\mu S\;\pd_\mu S-\frac{\yx}{2}(\bar\chic\chi+\bar\chi\chic)S
	\to
	\frac{i}{2}\bar\psi\pds\psi+\frac{1}{2}\pd^\mu\phi_S\;\pd_\mu\phi_S
	-\frac{\yx\vs}{2}\bar\psi\psi-\frac{\yx}{2}\bar\psi\psi\phi_S
	\eeq
where $\psi=\psi^\conj\equiv\chi+\chic$ is a Majorana mass eigenstate with $m_\psi=\yx\vs$. 

Here, as in the models discussed earlier,  there are two physical (mass eigenstates) scalar degrees of freedom, $h_1$ and $h_2$, that are linear combinations of $\phi_H$ and $\phi_S$.
Note that because of appropriate normalization of terms involving $S$ in the potential (\ref{pot_fdm}) the mass matrix and its diagonalization remain the same as in the other models. It is convenient to use the analogous input parameters to discuss this model: $\vs$, 
$\sin\alpha$, $m_2$, and $\mdm = m_\psi$.

Positivity and minimization conditions for this model will be discussed in section~\ref{theo_con}.

\begin{figure}[H]
	\centering
	\includegraphics[height=0.2\textwidth]{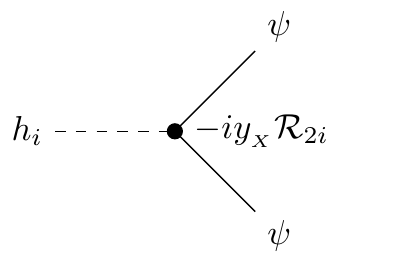}
	\caption{The vertex relevant for the FDM model.}
	\label{fig:FDM_ver}
	\end{figure}

Figure~\ref{fig:FDM_ver} presents the vertex relevant for the calculation of annihilation cross section in the FDM model.

\section{Astrophysical and other constraints}
\label{astro_con}

Hereafter we are going to allow for resonant DM annihilation process, so we will adopt 
the Breit-Wigner propagators for mediating particles, i.e. the Higgs bosons $h_{1,2}$. $\Gamma_{1,2}$
will denote the total width of $h_{1,2}$, respectively.

\subsection{Dark matter abundance}

The thermally averaged cross section for DM annihilation into a SM fermion-anti-fermion pair, $\sigma(\dm\;\dm \to \bar f f)$, reads:\footnote{Other final states are not accessible kinematically for mass ranges considered here.}
	\beq\label{ann_xsec}
	\begin{aligned}
	\langle\sigma v\rangle&=\frac{n_c}{3}\frac{m_\dm m_f^2}{\pi} \cdot \X \cdot
		\frac{\left(m_\dm^2-m_f^2\right)^{3/2}}{\left[(4m_\dm^2-m_1^2)^2+m_1^2\Gamma_1^2\right]\left[(4m_\dm^2-m_2^2)^2+m_2^2\Gamma_2^2\right]}\cdot\\
	&\quad\times
	\begin{cases}
		12+\O\left[\left(\frac{m_\dm}{T}\right)^{-1}\right] & (\sdm)\\
		1+\O\left[\left(\frac{m_\dm}{T}\right)^{-1}\right] & (\vdm)\\
		\frac94\left(\frac{m_\dm}{T}\right)^{-1}+\O\left[\left(\frac{m_\dm}{T}\right)^{-2}\right] & (\fdm)
	\end{cases}\;,
	\end{aligned}
	\eeq
with $n_c=1 (3)$ for $f$ being lepton (quark) and the variable $\X$ defined\footnote{Note that at the tree level $\X$ reduces to $\kappa^2$.} as
	\beq
	\qquad \X \equiv (\sin\alpha\cos\alpha)^2
	\frac{\left[(m_1^2-m_2^2)^2+(m_1\Gamma_1-m_2\Gamma_2)^2\right]}{v^2\vs^2}\;.
	\label{X_def}
	\eeq
The DM abundance observed by the Planck Collaboration \cite{Ade:2015xua},
\beq
\left(\Omega h^2\right)^\text{obs}_\dm= 0.1186\pm0.002\;,
\label{Omega}
\eeq
constraints the annihilation cross section at the freeze-out temperature by 
	\beq
	\langle\sigma v\rangle\at{\text{freeze out}}=
	(n+1)\cdot2.2\cdot 10^{-26}\text{ cm}^3\,\text{s}^{-1}=
	(n+1)\cdot1.9\cdot 10^{-9}\gev^{-2}\;,
	\label{DD_lim}
	\eeq
what corresponds to the current value of annihilation cross section equal to
	\bal
	&\langle\sigma v\rangle\at{\text{now}}=
	(T_0/T_f)^n\cdot\langle\sigma v\rangle\at{\text{freeze out}}=
	(T_0/T_f)^n\cdot(n+1)\cdot1.9\cdot 10^{-9}\gev^{-2}\;,
	\label{id_xsec}
	\eal
where $T_0$ is the present CMB temperature while $T_f\sim m_\dm/25$ is temperature at the moment of freeze out. 
Value of $n$ is 0 for the bosonic models (pGDM, VDM) and 1 for the FDM.

Hence, keeping only the leading ($b\bar b$) contribution in eq.~(\ref{ann_xsec}), we obtain the following constraint
\beq
	\begin{aligned}\label{x_omega}
	\X&\simeq
	\frac{\left[(m_1^2-4m_\dm^2)^2+m_1^2\Gamma_1^2\right]\,
	\left[(m_2^2-4m_\dm^2)^2+m_2^2\Gamma_2^2\right]}{m_\dm(m_\dm^2-m_b^2)^{3/2}}\times\\
	&\times 3.5 \cdot 10^{-10}\gev^{-4}\cdot
	\begin{cases}
	1/12&(\sdm)\\
		1&(\vdm)\\
		22&(\fdm)
	\end{cases}\,.
	\end{aligned}
\eeq

\subsection{Dark matter indirect detection}

Since we fix the DM abundance to its observed value (\ref{Omega}), the present annihilation cross section is also fixed by (\ref{id_xsec}), so that it remains to be a function of $\mdm$ only. Therefore, the limit on the present annihilation cross section, for instance from Fermi-LAT~\cite{Fermi-LAT:2016uux}, implies a lower limit 
on DM mass. For the bosonic models (pGDM, VDM),  adopting data for the $b\bar b$ final state, 
one obtains $m_\dm \gsim 20\gev$. Hereafter, we will consider this region only.
In the case of the FDM, the extra suppression by $T_0/T_f$ implies that the cross section is by a factor of $10^{-11}$--$10^{-13}$ smaller than for the bosonic models and, therefore, there is no constraint on $m_\psi$. 

\subsection{Dark matter direct detection}
\label{DD_dd_vdm}

\begin{figure}[!ht]\begin{center}
	\includegraphics[width=0.46\textwidth]{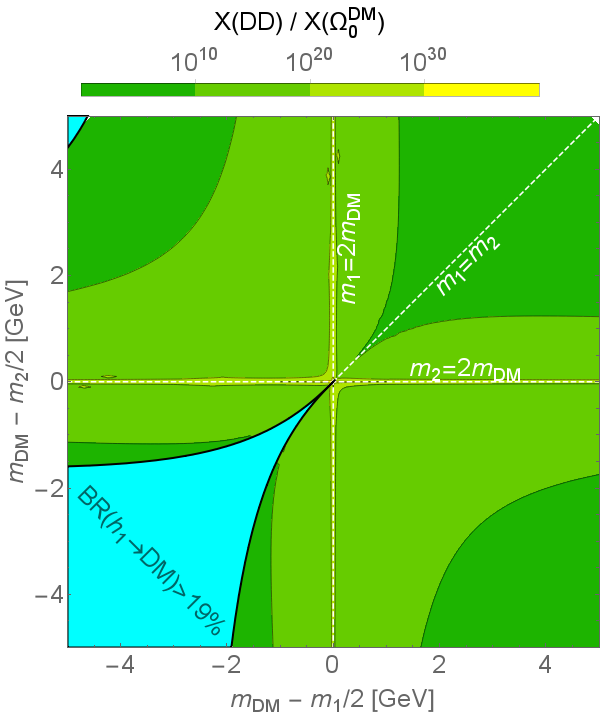}
	\caption{Comparison between the DD upper bound for the value of $\X$ (denoted by $\X(\text{DD})$, see eq.~(\ref{x_dd})) and the value providing correct relic density (denoted by $\X(\Omega_0^\dm)$, see eq.~(\ref{x_omega})), in the case of the pGDM. Since the upper bound is always higher than the required value, the DD constraint does not affect the range of ($m_2$,$m_\dm$) in the case of this model.}
	\label{fig:SDMConstraintsComparison}
	\end{center}
	\end{figure}

The DM direct detection (DD) experiments impose severe constraints on the parameter space of DM models.
In the models discussed here the spin-independent cross sections for the DM-nucleon scattering are given by 
\beq
	\begin{aligned}
	\sigma_\text{SI}& \simeq \frac{\mu^2 f_N^2}{\pi}\cdot \X \cdot\frac{m_\dm^2 m_N^2}{m_1^4m_2^4}
	\begin{cases}
		\left[\frac{\mathcal{A}}{64\pi^2 v\vs^2}\right]^2 & (\sdm)\\
		1&(\vdm), (\fdm)\,,
	\end{cases}
	\end{aligned}
	\label{SI_cross}
\eeq
where $m_N$ denotes nucleon mass and $\mu$ is the reduced mass for the DM-nucleon system while for the form factor we have adopted $f_N\simeq 0.3\gev$.
Widths and momentum transfer in the denominator have been neglected as much smaller than masses. It turns out that in the pGDM model the cross section vanishes~\cite{Azevedo:2018oxv,Azevedo:2018exj,Gross:2017dan} in the limit of zero momentum transfer, so 1-loop calculations are needed. The 1-loop results are encoded above through the factor containing ${\cal A}$, defined according to \cite{Azevedo:2018exj}\footnote{In appendix B of \cite{Azevedo:2018exj}, the factor $1/(2\pi)^4$ in definitions of loop integrals should be replaced  by $1/(i\pi^2)$. Nonetheless, all results in the main text of the paper are correct.} as
	\beq
	\begin{aligned}
	\mathcal{A} = &a_1\cdot C(0,m_\dm;m_1,m_2,m_\dm)+\\
	&a_2\cdot D(0,0,m_\dm;m_1,m_1,m_2,m_\dm)+\\
	&a_3\cdot D(0,0,m_\dm;m_1,m_2,m_2,m_\dm)
	\end{aligned}
	\label{loop_res}
	\eeq
with
	\beq
	\begin{aligned}
	a_1&=4(m_1^2\sin^2\alpha+m_2^2\cos^2\alpha)
		\left[2v(m_1^2\sin^2\alpha+m_2^2\cos^2\alpha)-(m_1^2-m_2^2)\vs\sin2\alpha\right]\;,\\
	a_2&=-2m_1^4\sin\alpha
		\left[(m_1^2+5m_2^2)\vs\cos\alpha-(m_1^2-m_2^2)(\vs\cos3\alpha+4v\sin^3\alpha)\right]\;,\\
	a_3&=2m_2^4\cos\alpha
		\left[(5m_1^2+m_2^2)\vs\sin\alpha-(m_1^2-m_2^2)(\vs\sin3\alpha+4v\cos^3\alpha)\right]\;,
	\end{aligned}
	\label{loop_coeffs}
	\eeq
where the functions $C$ and $D$ are defined in appendix~\ref{sec:pave}. In eq.~(\ref{loop_coeffs}), the sign of $\sin\alpha$ is relevant, what seems to contradict our statement that chosing $\sin\alpha>0$ does not spoil generality of our considerations. However, the only place where we use results of eq.~(\ref{SI_cross}) for the pGDM is the comparison in figure~\ref{fig:SDMConstraintsComparison}. Regardless of the sign of $\sin\alpha$, the conclusion that $\X(\text{DD})$ is orders of magnitude larger than $\X(\Omega_0^\dm)$ remains true, and hence, we do not have to consider the $\sin\alpha<0$ case separately.

For practical purposes, the XENON1T limit~\cite{Aprile:2018dbl} for $m_\dm\gtrsim40\gev$ can be parametrized as follows
	\beq
	\frac{\sigma_\text{SI}^{\max}}{1\text{ cm}^2}
	\simeq \frac{m_\dm}{1\gev}\cdot10^{-48.05}\;.
	\eeq
Hence, $\X$ is constrained from above by DD limit:
	\bea
	\X &\lsim &
	\frac{m_1^4 m_2^4}{m_\dm m_N^2 f_N^2}\frac{\pi}{\mu^2}
	\frac{1\text{ cm}^2}{1\gev}\cdot10^{-48.05}
	\begin{cases}
		\left[\frac{\mathcal{A}}{64\pi^2 v\vs^2}\right]^{-2} & (\sdm)\\
		1&(\vdm), (\fdm)\,,
	\end{cases}\non\\
	&\simeq &
	\frac{m_2^4}{m_\dm}\cdot 2.5 \cdot 10^{-11}\gev^{-3}
	\begin{cases}
		\left[\frac{\mathcal{A}}{64\pi^2 v\vs^2}\right]^{-2} & (\sdm)\\
		1&(\vdm), (\fdm)
	\end{cases}\;.\label{x_dd}
	\eea

It turns out that in the considered range of parameters, in the case of the pGDM, the DD upper bound on the value of $\X$ is always higher than the value corresponding to the correct relic density, see figure~\ref{fig:SDMConstraintsComparison}. 
Therefore, in the case of the pGDM, the DD constraint does not limit the range of $(m_2,m_\dm)$.
	\pagebreak
%
\subsection{Collider constraints}
\label{coll_con}

The mixing angle $\alpha$ is constrained from the measurement of the SM signal strength $\mu_\text{LHC}$. 
The latest LHC bound is $\mu_\text{LHC}=1.09\pm 0.11$  which amounts to $\sin^2\alpha < 0.13$ at the $2\sigma$ CL~\cite{Chang:2017ynj}. Hereafter we will adopt a bit stronger limit $\sin\alpha<0.30$.

When the DM mass is smaller than half of the SM-like Higgs boson $h_1$, $\mdm < m_1/2$, the Higgs invisible decay 
provides another constraint on DM scenarios. In the models discussed here, the width for invisible decays are as follows
	\bal
	\Gamma_{h_i\to\dm}&=\frac{\R_{2i}^2}{\vs^2}\cdot
	\frac{m_i^3}{32\pi}\sqrt{1-\frac{4m_\dm^2}{m_i^2}}\times\notag\\
		&\quad\times\begin{cases}
		1 & (\sdm)\\
		1 - 4\frac{m_\dm^2}{m_i^2} + 12 \left(\frac{m_\dm^2}{m_i^2}\right)^2 & (\vdm)\\
		2\frac{m_\dm^2}{m_i^2}\left(1-4\frac{m_\dm^2}{m_i^2}\right) & (\fdm)
		\end{cases}\;.
	\eal
Current LHC measurements \cite{Sirunyan:2018owy} provides the following limit on the invisible branchig ratio:
	\beq
	\br(h_1\to \text{inv}) < 19\%
	\label{inv_br}
	\eeq
at the 95\% CL.

\subsection{Theoretical constraints}
\label{theo_con}

In order to ensure that the leading order calculations adopted here are meaningful, we impose the following perturbativity conditions on the $\uone$ gauge coupling in the VDM model and the Yukawa coupling in the FDM model:
 $\gx<4\pi$ and $\yx<4\pi$. Both of them correspond to $\vs>\frac{\mdm}{4\pi}$.
In the pGDM model, the $AAh_i$ coupling is proportional to $m_i^2/\vs$ (cf. figure~\ref{SDM_ver}), therefore we also require $m_i/\vs<4\pi$ $(i=1,2)$. It is interesting to note that there exist regions (e.g. $m_2 \sim m_1$) in the parameter space where the proper abundance of DM requires small $\vs$. In these regions some quartic couplings might be too large (non-perturbative), since $\lambda_S \propto m_i^2/\vs^2$ and $\kappa \propto (m_1^2-m_2^2)/(v\vs)$, see figure~\ref{fig:cons_tot}. Therefore we also impose the conditions: $\lambda_S, |\kappa| < 4\pi$. Summing up, the conditions adopted here in order to ensure perturbativity within considered models are
	\begin{align}\label{eq:pertCond}
	&&\underbrace{\frac{m_\dm}{\vs}<4\pi}_{\text{for VDM and FDM}}\;,
	&&\underbrace{\frac{m_i}{\vs}<4\pi}_{\text{for pGDM}}\;,
	&&\lambda_S<4\pi\;,
	&&|\kappa|<4\pi\;.
	\end{align}

Let us now consider conditions for stability of the vacuum state. Scalar potentials of the models read (see eqs.~(\ref{pot_sdm}), (\ref{pot_vdm}) and (\ref{pot_fdm})):
	\begin{align}
	V_\text{pGDM}(H,S)&=
	-\mu_H^2|H|^2+\lambda_H|H|^4-\mu_S^2|S|^2+\lambda_S|S|^4+\kappa|H|^2|S|^2+\\
	\notag
	&\quad+\mu^2(S^2+{S^*}^2)\;,\\
	V_\text{VDM}(H,S)&=
	-\mu_H^2|H|^2+\lambda_H|H|^4-\mu_S^2|S|^2+\lambda_S|S|^4
	+\kappa|H|^2|S|^2\;,\\
	V_\text{FDM}(H,S)&=
	-\mu_H^2|H|^2+\lambda_H|H|^4-\frac{\mu_S^2}{2}S^2+\frac{\lambda_S}{4}S^4
	+\frac{\kappa}{2}|H|^2S^2\;.
	\end{align}

To ensure asymptotic positivity of all the potentials, the following conditions must be satisfied:
	\begin{align}\label{eq:asympPos}
	\lambda_H&>0\;,\qquad\lambda_S>0\;,\qquad
	\kappa>-2\sqrt{\lambda_H\lambda_S}\;.
	\end{align}

Vacuum expectation values of the scalar fields $H$, $S$ are denoted  as follows 
	\begin{align}
	\langle H \rangle &= v/\sqrt2\,,&	\langle S \rangle &= \vs/\sqrt2&	&(\text{pGDM, VDM})\;,\\
	\langle H \rangle &= v/\sqrt2\,,&	\langle S \rangle &= \vs&			&(\text{FDM})\;.
	\end{align}

In each case, $v,\vs\neq 0$ must minimize the value of the potential. The corresponding point in the $(H,S)$ space is a critical one if and only if
	\begin{align}\label{eq:vevs-pGDM}
	v^2&=2\,\frac{2\lambda_S\mu_H^2-\kappa(\mu_S^2-2\mu^2)}{4\lambda_H\lambda_S-\kappa^2}\;,&
	\vs^2&=2\,\frac{2\lambda_H(\mu_S^2-2\mu^2)-\kappa\mu_H^2}{4\lambda_H\lambda_S-\kappa^2}
	\end{align}
in the case of the pGDM and
	\begin{align}\label{eq:vevs-VFDM}
	v^2&=2\,\frac{2\lambda_S\mu_H^2-\kappa\mu_S^2}{4\lambda_H\lambda_S-\kappa^2}\;,&
	\vs^2&=2\,\frac{2\lambda_H\mu_S^2-\kappa\mu_H^2}{4\lambda_H\lambda_S-\kappa^2}
	\end{align}
in the case of the VDM and the FDM.

To ensure that the critical point is a strict minimum, we demand the second derivative of the potential to be positive definite, therefore
	\begin{align}
	0&<\partial^2_{H,H}V=4\lambda_Hv^2\;,\\
	0&<\det(D^2V)
	=4v^2\vs^2(4\lambda_H\lambda_S-\kappa^2)\cdot
	\begin{cases}
	1	&(\text{pGDM, VDM})\\
	1/2	&(\text{FDM})
	\end{cases}\;.
	\end{align}
Hence, assuming $v^2$, $\vs^2$ and $\lambda_H$ are positive, the following condition must hold
	\begin{align}\label{eq:sctrictMin}
	4\lambda_H\lambda_S-\kappa^2>0\;.
	\end{align}

Positivity of the vevs squared requires (cf. eqs.~(\ref{eq:vevs-pGDM}) and (\ref{eq:vevs-VFDM}))
	\begin{align}\label{eq:vevsPos}
	2\lambda_S\mu_H^2-\kappa(\mu_S^2-2\mu^2)&>0\;,&
	2\lambda_H(\mu_S^2-2\mu^2)-\kappa\mu_H^2&>0
	\end{align}
in the case of the pGDM and
	\begin{align}
	2\lambda_S\mu_H^2-\kappa\mu_S^2&>0\;,&
	2\lambda_H\mu_S^2-\kappa\mu_H^2&>0
	\end{align}
in the case of the VDM and the FDM.

Let us check when the points given by eqs.~(\ref{eq:vevs-pGDM}) and (\ref{eq:vevs-VFDM}) are global minima. In the case of the pGDM, due to the presence of the $\mu^2(S^2+S^{*2})$ term, in principle the phase of the vacuum expectation value of $S$ could be relevant. Hence, let us assume that $\langle S\rangle=(\vs+i\va)/\sqrt2$. Now, we have to minimize the potential with respect to $v$, $\vs$ and $\va$. There are six critical points of the potential, namely
	\begin{align}
	(v^2\,,\;\vs^2\,,\;\va^2)&=(0\,,\;0\,,\;0)\;,&
	V&=0\;,\\
	\label{eq:min1pGDM}
	(v^2\,,\;\vs^2\,,\;\va^2)&=\left(\frac{\mu_H^2}{2\lambda_H}\,,\;0\,,\;0\right)\;,&
	V&=-\frac{\mu_H^4}{4\lambda_H}\;,\\
	\label{eq:min2pGDM}
	(v^2\,,\;\vs^2\,,\;\va^2)&=\left(0\,,\;\frac{\mu_S^2-2\mu^2}{2\lambda_S}\,,\;0\right)\;,&
	V&=-\frac{(\mu_S^2-2\mu^2)^2}{4\lambda_S}\;,\\
	\label{eq:min3pGDM}
	(v^2\,,\;\vs^2\,,\;\va^2)&=\left(0\,,\;0\,,\;\frac{\mu_S^2+2\mu^2}{2\lambda_S}\right)\;,&
	V&=-\frac{(\mu_S^2+2\mu^2)^2}{4\lambda_S}\;,
	\end{align}
	\vspace{-0.7cm}
	\begin{align}
	\notag
	(v^2\,,\;\vs^2\,,\;\va^2)&=\left(2\,\frac{2\lambda_S\mu_H^2-\kappa(\mu_S^2+2\mu^2)}
	{4\lambda_H\lambda_S-\kappa^2}\,,\;
	0\,,\;
	2\,\frac{2\lambda_H(\mu_S^2+2\mu^2)-\kappa\mu_H^2}{4\lambda_H\lambda_S-\kappa^2}\right)\;,\\
	\label{eq:min4pGDM}
	V&=-\frac{\lambda_H(\mu_S^2+2\mu^2)^2+\lambda_S\mu_H^4-\kappa\mu_H^2(\mu_S^2+2\mu^2)}
	{4\lambda_H\lambda_S-\kappa^2}\;,\\
	\notag
	(v^2\,,\;\vs^2\,,\;\va^2)&=\left(2\,\frac{2\lambda_S\mu_H^2-\kappa(\mu_S^2-2\mu^2)}
	{4\lambda_H\lambda_S-\kappa^2}\,,\;
	2\,\frac{2\lambda_H(\mu_S^2-2\mu^2)-\kappa\mu_H^2}{4\lambda_H\lambda_S-\kappa^2}\,,\;
	0\right)\;,\\
	\label{eq:min5pGDM}
	V&=-\frac{\lambda_H(\mu_S^2-2\mu^2)^2+\lambda_S\mu_H^4-\kappa\mu_H^2(\mu_S^2-2\mu^2)}
	{4\lambda_H\lambda_S-\kappa^2}\;.
	\end{align}
Assuming the asymptotic positivity conditions (\ref{eq:asympPos}), the strict-minimum condition (\ref{eq:sctrictMin}) and positivity of $m_\dm^2=-4\mu^2$, minimum (\ref{eq:min5pGDM}) is always smaller than (\ref{eq:min1pGDM}) and (\ref{eq:min2pGDM}). To ensure that minimum (\ref{eq:min5pGDM}) is smaller than (\ref{eq:min4pGDM}), the following additional condition must hold:
	\begin{align}
	\label{eq:pGDMglobalMinCond1}
	2\lambda_H\mu_S^2-\kappa\mu_H^2>0\;.
	\end{align}
Value of (\ref{eq:min3pGDM}) is obviously greater than (\ref{eq:min2pGDM}) and, therefore, greater than (\ref{eq:min5pGDM}) if
	\begin{align}
	\label{eq:pGDMglobalMinCond2}
	\mu_S^2>0\;.
	\end{align}
Both of these conditions are checked for considered region of parameter space at the end of this subsection.

In the case of the VDM and the FDM, we can assume that $\langle S\rangle$ is purely real without losing generality. Therefore,  $V$ is minimized with respect to $v$ and $\vs$. The critical points are

	\begin{minipage}{\textwidth}
	\begin{align}
	(v^2\,,\;\vs^2)&=(0\,,\;0)\;,&
	V&=0\;,\\
	\label{eq:min1VFDM}
	(v^2\,,\;\vs^2)&=\left(\frac{\mu_H^2}{2\lambda_H}\,,\;0\right)\;,&
	V&=-\frac{\mu_H^4}{4\lambda_H}\;,\\
	\label{eq:min2VFDM}
	(v^2\,,\;\vs^2)&=\left(0\,,\;\frac{\mu_S^2}{2\lambda_S}\right)\;,&
	V&=-\frac{\mu_S^4}{4\lambda_S}\;,
	\end{align}
	\vspace{-0.7cm}
	\begin{align}\begin{aligned}
	(v^2\,,\;\vs^2)&=\left(2\,\frac{2\lambda_S\mu_H^2-\kappa\mu_S^2}{4\lambda_H\lambda_S-\kappa^2}\,,\;
	2\,\frac{2\lambda_H\mu_S^2-\kappa\mu_H^2}{4\lambda_H\lambda_S-\kappa^2}\right)\;,\\
	\label{eq:min3VFDM}
	V&=-\frac{\lambda_H\mu_S^4+\lambda_S\mu_H^4-\kappa\mu_H^2\mu_S^2}{4\lambda_H\lambda_S-\kappa^2}\;.
	\end{aligned}\end{align}\\
	\end{minipage}

This time, the asymptotic positivity conditions (\ref{eq:asympPos}) and the strict-minimum condition (\ref{eq:sctrictMin}) are enough to keep (\ref{eq:min3VFDM}) a global minimum.

We can express parameters of the potential in terms of the input parameters: $m_1$, $m_2$, $v$, $\vs$, $m_\dm$ and $\sin\alpha$ as follows:
	\begin{gather}
	\label{relation_s}
	\mu^2=-\frac{1}{4}m_\dm^2\text{ (pGDM case only)}\;,\qquad
	\kappa=\frac{(m_1^2-m_2^2)\sin2\alpha}{2v\vs}\;,\\
	\mu_H^2=\frac{1}{2}m_1^2\cos^2\alpha+\frac{1}{2}m_2^2\sin^2\alpha
	+\frac{1}{4}\frac{\vs}{v}(m_1^2-m_2^2)\sin2\alpha\;,\\
	\begin{aligned}
	\mu_S^2&=\frac{1}{2}m_1^2\sin^2\alpha+\frac{1}{2}m_2^2\cos^2\alpha
	+\frac{1}{4}\frac{v}{\vs}(m_1^2-m_2^2)\sin2\alpha+\\
	&\quad
	-\begin{cases}
	\frac12m_\dm^2	&\text{(pGDM)}\\
	0				&\text{(VDM, FDM)}
	\end{cases}\;,
	\end{aligned}\\
	\lambda_H=\frac{m_1^2\cos^2\alpha+m_2^2\sin^2\alpha}{2v^2}\;,\qquad
	\lambda_S=\frac{m_1^2\sin^2\alpha+m_2^2\cos^2\alpha}{2\vs^2}\;.
	\end{gather}

It appears that the stability and positivity conditions (\ref{eq:asympPos}), (\ref{eq:sctrictMin}) and (\ref{eq:vevsPos}) expressed in terms of the input parameters are automatically satisfied:

	\begin{minipage}{\textwidth}
	\begin{align}
	0&<\lambda_H&
	&\Leftrightarrow&
	0&<\frac{m_1^2\cos^2\alpha+m_2^2\sin^2\alpha}{2v^2}\;,\\
	0&<\lambda_S&
	&\Leftrightarrow&
	0&<\frac{m_1^2\sin^2\alpha+m_2^2\cos^2\alpha}{2\vs^2}\;,\\
	0&<4\lambda_H\lambda_S-\kappa^2&
	&\Leftrightarrow&
	0&<\frac{m_1^2m_2^2}{v^2\vs^2}\;,
	\end{align}
	\vspace{-0.7cm}
	\begin{align}
	0&<\begin{cases}
	2\lambda_S\mu_H^2-\kappa(\mu_S^2-2\mu^2)	&\text{(pGDM)}\\
	2\lambda_S\mu_H^2-\kappa\mu_S^2				&\text{(VDM, FDM)}\\
	\end{cases}&
	&\Leftrightarrow&
	0&<\frac{m_1^2m_2^2}{2\vs^2}\;,\\
	0&<\begin{cases}
	2\lambda_H(\mu_S^2-2\mu^2)-\kappa\mu_H^2	&\text{(pGDM)}\\
	2\lambda_H\mu_S^2-\kappa\mu_H^2				&\text{(VDM, FDM)}\\
	\end{cases}&
	&\Leftrightarrow&
	0&<\frac{m_1^2m_2^2}{2v^2}\;.
	\end{align}\\
	\end{minipage}

In fact, our choice of the input set implicitly assumes that coefficients of $V$ are such that $v^2,\vs^2,m_1^2,m_2^2>0$.

The global-minimum conditions (\ref{eq:pGDMglobalMinCond1}) and (\ref{eq:pGDMglobalMinCond2}) for the case of the pGDM are expressed in terms of the input parameters as follows:
	\begin{gather}
	\label{eq:pGDMglobalMinCondPhysVars1}
	0<(2\lambda_H\mu_S^2-\kappa\mu_H^2)
	\Leftrightarrow
	0<\frac{2m_1^2m_2^2-(m_1^2+m_2^2)m_\dm^2+(-m_1^2+m_2^2)m_A^2\cos(2\alpha)}{4 v^2}\;,\\
	\label{eq:pGDMglobalMinCondPhysVars2}
	0<\mu_S^2
	\quad\Leftrightarrow\quad
	0<\frac{1}{2}m_1^2\sin^2\alpha+\frac{1}{2}m_2^2\cos^2\alpha
	+\frac{1}{4}\frac{v}{\vs}(m_1^2-m_2^2)\sin2\alpha-\frac12m_\dm^2\;.
	\end{gather}

\begin{figure}[!ht]
	\centering
	\includegraphics[width=0.45\textwidth]{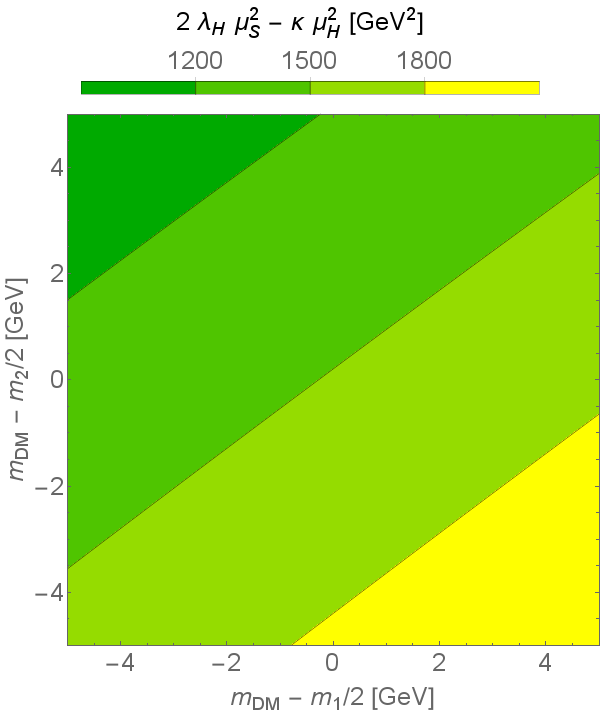}
	\includegraphics[width=0.45\textwidth]{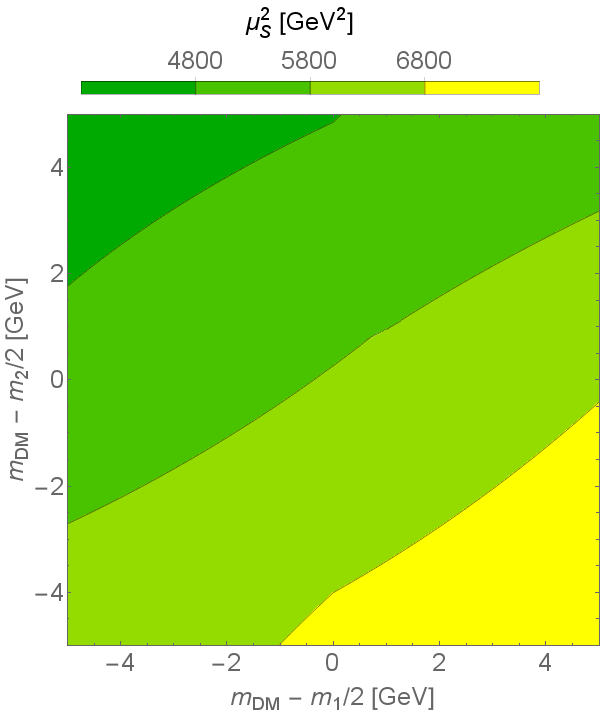}
	\caption{Numerical test of the conditions (\ref{eq:pGDMglobalMinCondPhysVars1}) (left pannel) and (\ref{eq:pGDMglobalMinCondPhysVars2}) (right pannel) for globalness of the minimum of the scalar potential in the pGDM model. If the plotted values are positive, the conditions are satisfied. Value of $\sin\alpha$ has been assumed to be $0.3$. Value of $\vs$ has been calculated from eq.~(\ref{x_omega}).}
	\label{fig:globalMinCond}
\end{figure}

It can be numerically shown (see figure~\ref{fig:globalMinCond}) that in the considered range of masses these conditions are always satisfied.

\section{Production of DM pairs at future \boldmath{$e^+e^-$} colliders}
\label{Production}

\begin{figure}[!ht]
	\centering
	\includegraphics[width=0.45\textwidth]{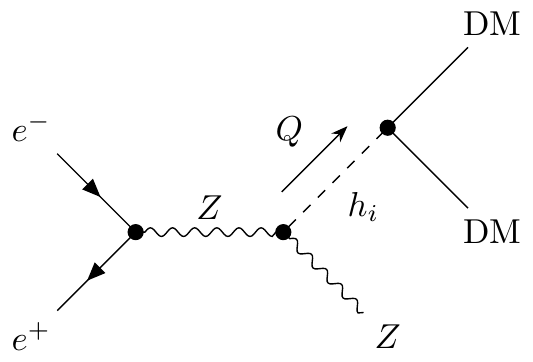}
	\caption{Feynman diagram for the considered channel of DM production. In the diagram, DM denotes the dark particle that is either $A$, $X$ or $\psi$.}
	\label{fig:eeHZ}
\end{figure}

The DM models can be tested at $e^+e^-$ collider experiments. In particular, these experiments allow  for the copious production of DM states associated with a $Z$~boson, what is referred to as so called Higgsstrahlung process or mono-$Z$ emission \cite{Dreiner:2012xm,Yu:2014ula,Neng:2014mga,Ko:2016xwd,Liu:2017lpo,Rawat:2017fak,Kamon:2017yfx}, see figure~\ref{fig:eeHZ}.
We assume that the energy of the $Z$ boson can be reconstructed from data, therefore allowing for determination of the recoil mass ($\sqrt{Q^2}$), corresponding to the invariant mass of the dark particles. 

The differential cross section for DM pair production at $e^+e^-$ colliders reads 
	\bal
	\frac{d\sigma}{dQ^2}&=
	\frac{\sigma_\sm(s,Q^2)\,v^2}{32\pi^2}\frac{\X\cdot \left(Q^2\right)^2}
	{\left[(Q^2-m_1^2)^2+(m_1\Gamma_1)^2\right]\left[(Q^2-m_2^2)^2+(m_2\Gamma_2)^2\right]}\times\non\\
	&\times\sqrt{1-4\frac{m_\dm^2}{Q^2}}\cdot
		\begin{cases}
		1 & (\sdm) \\
		1 - 4\frac{m_\dm^2}{Q^2} + 12 \left(\frac{m_\dm^2}{Q^2}\right)^2 & (\vdm) \\
		2\frac{m_\dm^2}{Q^2}\left(1-4\frac{m_\dm^2}{Q^2}\right) & (\fdm) 
		\end{cases}\;,
	\label{d_sig}
	\eal
where the parameter $\X$ is defined in eq.~(\ref{X_def}) and
	\bal\begin{aligned}
	\sigma_\sm(s,Q^2)&\equiv\frac{g_V^2+g_A^2}{24\pi}
	\left(\frac{g^2}{\cos\theta_W^2}\frac{1}{s-m_Z^2}\right)^2\times\\
	&\times\frac{\lambda^{1/2}(s,Q^2,m_Z^2)\left[12\,s\,m_Z^2+\lambda(s,Q^2,m_Z^2)\right]}{8s^2}
	\end{aligned}\eal
is the cross section for the $e^+e^-\to Zh_\sm$ process, with mass of the SM Higgs particle equal to $\sqrt{Q^2}$. Here, $\lambda(a,b,c)$ denotes the K\"all\'en function, defined in appendix~\ref{propagator}, and $g_V$, $g_A$ stand for the vector and axial coupling, respectively.\footnote{In the case of polarized beams, ${g_V^2+g_A^2}$ factor has to be replaced with ${(1-P_+P_-)(g_V^2+g_A^2)+2g_Vg_A(P_+-P_-)}$, where $P_\pm$ denotes polarizations of $e^\pm$ beams.} The above result has been obtained by adopting the standard Breit-Wigner propagators for the virtual/real Higgs bosons $h_i$.  

Note that in the limit of $m_2 \to m_1$ the cross section $d\sigma/dQ^2$ (\ref{d_sig}) seems to be amplified for $h_i$ being on-shell, i.e. $Q^2\to m_{1,2}^2$. This is a surprising observation since, on the other hand,    
the second relation in (\ref{relation_s}) between masses and the portal coupling $\kappa$ implies that in the limit $m_2 \to m_1$, whenever $\vs \neq 0$, $\kappa \to 0$ so that the dark sector decouples in each model discussed here. Therefore, all cross sections for DM production or annihilation from the SM must vanish in this limit. Behaviour of the cross sections in this limit is potentially important phenomenologically, therefore in the following we are going to investigate the $Q^2\to m_{1,2}^2$ limit in more details.

\begin{figure}[!ht]\begin{center}
	\includegraphics[width=0.3\textwidth]{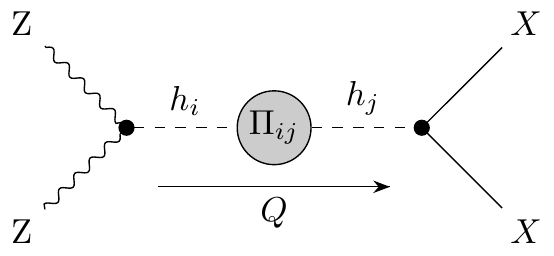}
	\caption{The Higgs-boson mediators with their self-energies.}
	\label{fig:prop}
	\end{center}\end{figure}

Let's investigate the parameter $\X$. First, it is easy to see that 
\beq
\lim_{m_2 \to m_1}\X= 
\left[\sin 2\alpha \; \frac{m_1\; (\Gamma_1-\Gamma_2)}{2 v\vs}\right]^2\,.
\label{X_lim}
\eeq
From (\ref{relation_s}) one finds that if $\vs\neq 0$ then the limit $m_2 \to m_1$ implies 
$\kappa \to 0$ and $\lambda_H v^2-\lambda_S\vs^2 \to 0$. Therefore,  according to (\ref{mix_alp}) $\tan 2\alpha$
is undefined. For instance, for fixed $\lambda_H$, $v$ and $\vs$ it is easy to see that, approaching the limiting point $(\lambda_H(v/\vs)^2,0)$ in the $(\lambda_S,\kappa)$ plane, one can get $\alpha=0$, $\alpha=\pi/4$ or $\alpha=1/2 \arctan (v/\vs)$, choosing the corresponding trajectories: $\kappa=0$, $\lambda_S=\lambda_H(v/\vs)^2$ or $\kappa=-\lambda_S+\lambda_H(v/\vs)^2$, respectively. 
Since in the limit $m_2 \to m_1$ neither $\sin 2\alpha \to 0$ nor $\Gamma_1\to \Gamma_2$, so $\X$ does not vanish, in spite of justified arguments mentioned above. The solution to this puzzle lies in the fact that for $m_2 \to m_1$ also off-diagonal ($i\neq j$) Higgs boson self-energies (see figure~\ref{fig:prop}) are relevant and should be resummed so the naive, diagonal, Breit-Wigner propagators are not appropriate. To illustrate this point let us consider the $e^+e^-\to ZXX$ process. The matrix element reads
	\beq\begin{aligned}
	\M&=\M_{e^+e^-\to Zh_i}(Q^2)\cdot\Delta_{ij}(Q^2)\cdot\M_{h_j\to XX}(Q^2)=\\
	&=\M_{e^+e^-\to Zh}(Q^2)\cdot
	\underbrace{\R_{1i}\cdot\Delta_{ij}(Q^2)\cdot\R_{2j}}_{\what\Delta(Q^2)}
	\cdot\M_{h\to XX}(Q^2)\,.
	\end{aligned}\eeq

By $\what\Delta$ we denote the propagator contracted with the mixing matrix. From \cite{Duch:2018ucs} (see also 
\cite{Cacciapaglia:2009ic}), the contracted propagator can be expressed explicitly as
	\beq\begin{aligned}\label{eq:propFull}
	\what\Delta(Q^2)&=\R_{1i}\R_{2j}\cdot\frac{1}{\det D}
	{\overbrace{\Mdnd{Q^2-m_2^2+\Pi_{22}}{-\Pi_{12}}{-\Pi_{21}}{s-m_1^2+\Pi_{11}}}^{D}}_{ij}=\\
	&=\sin\alpha\cos\alpha\cdot
	\frac{(m_1^2-m_2^2)-(\Pi_{11}-\Pi_{22})+(\tan\alpha\cdot\Pi_{12}-\cot\alpha\cdot\Pi_{21})}
	{(Q^2-m_1^2+\Pi_{11})(Q^2-m_2^2+\Pi_{22})-\Pi_{12}\Pi_{21}}\;,
	\end{aligned}\eeq
where $\Pi_{ij}\equiv\Pi_{ij}(Q^2)$ denotes the imaginary part (multiplied by~$i$) of the $h_ih_j$ self energy, satisfying $\Pi_{ii}(m_i^2)=im_i\Gamma_i$. In magnitude, all of them are comparable to $m_i\Gamma_i$. Results for $\Pi_{ij}$ are collected in appendix~\ref{propagator}.

If $|m_1-m_2|\gg\Gamma_1,\Gamma_2$, then the first term in the denominator, ${(Q^2-m_1^2+\Pi_{11})(Q^2-m_2^2+\Pi_{22})}$, dominates for any $Q^2$, as well as the first term of the numerator, $(m_1^2-m_2^2)$. In such a case, the propagator can be approximated by
	\beq\label{eq:propBW}
	\what\Delta(Q^2)\simeq\sin\alpha\cos\alpha\cdot
	\frac{m_1^2-m_2^2}{(Q^2-m_1^2+\Pi_{11})(Q^2-m_2^2+\Pi_{22})}\,.
	\eeq
It is easy to see that the above propagator could be rewritten (dropping terms proportional to $\Gamma_{1,2}$ in the numerator) as 
	\beq
	\what\Delta(Q^2)\simeq\what\Delta^{(BW)}(Q^2)\equiv\sin\alpha\cos\alpha\cdot
	\left[\frac{1}{Q^2-m_1^2+im_1\Gamma_1} - \frac{1}{Q^2-m_2^2+im_2\Gamma_2}\right]\;,
	\label{BW_diff}
	\eeq
which reduces to the standard Breit-Wigner propagator. This simplified result has to be replaced by the full formula whenever $|m_1-m_2|$ is comparable to the widths. In order to investigate the case $m_1 \sim m_2$ one has to calculate $\Pi_{ij}$. The explicit calculation (see appendix~\ref{propagator}) confirms that
	\beq\begin{aligned}
	\left[\left(\Pi_{11}-\Pi_{22}\right)
	-\left(\tan\alpha\cdot\Pi_{12}-\cot\alpha\cdot\Pi_{21}\right)
	\right]\at{m_1=m_2}=0\,.
	\end{aligned}\eeq
Hence, the full propagator (\ref{eq:propFull}) vanishes in the limit $m_1=m_2$, exactly as it should. An important consequence of this result is that in the double-resonance region of $Q^2 \sim m_1^2 \sim m_2^2$, in the closest vicinity of $m_1=m_2$, the straightforward application of the Breit-Wigner strategy is not appropriate.

However, in practice, the region $|m_1-m_2| \lsim \Gamma_{1,2}$ is so narrow that the naive Breit-Wigner approximated resummation (\ref{BW_diff}) could be adopted, keeping in mind that exactly on the diagonal $m_1=m_2$ the cross sections do vanish.
 
\section{Constraints expected from future \boldmath{$e^+e^-$} colliders}
\label{coll_exp}

Production of the Standard Model Higgs boson in the Higgsstrahlung process is considered as a ``golden channel'' for  a model independent determination of the Higgs boson properties at future  $e^+e^-$ colliders.
By reconstructing the produced $Z$ bozon,  Higgsstrahlung events can be selected with high efficiency independently on the Higgs boson decay.

Largest sample of events can be selected when both leptonic and hadronic decay channels of the $Z$ boson are considered.
Reconstructing just the $Z$ boson is of particular interest when we look for rare processes involving the Higgs boson, for instance possible decays into DM states.
Events with mono-$Z$ production, and no other activity in the detector, can be considered as candidate events for the invisible Higgs boson decays, if the recoil mass, $\sqrt{Q^2}$, reconstructed from energy-momentum conservation, is consistent with the Higgs boson mass.
Highest sensitivity to invisible decays of the 125\,GeV boson is expected at $\sqrt{s} \simeq 250$\,GeV, corresponding to the maximum of the Higgsstrahlung cross section.
The main background processes that limit the sensitivity at this energy range are the production of $ZZ$ and $W^+W^-$ pairs, as well as single $Z$ production via the $WW$ fusion, $e^+ e^- \to \nu_e\bar{\nu}_e Z$. 
For the $Z$-boson pair production with one boson decaying into neutrinos, the final state reconstructed in the detector is
identical to the one expected for the invisible Higgs boson decays and the recoil mass can be significantly overestimated due to beams spectra\footnote{At linear $e^+e^-$ colliders the beamstrahlung effects result in the long tail in the beam energy spectra towards low energies. When the electron or positron participating in the collision has the initial energy much smaller than the nominal beam energy, the recoil mass can be significantly overestimated.} or large initial state radiation.
For hadronic $Z$-boson decays, also detector resolution effects, dominated by the jet energy resolution, are very important.  
Nevertheless, due to branching fraction much larger than in the leptonic case, the expected limits on the invisible decays of the 125\,GeV Higgs boson are dominated by hadronic $Z$ decay measurements.
For 2000\,fb$^{-1}$ of data collected at 250\,GeV ILC, the expected limit on the invisible branching fraction is 0.23\%, when combining hadronic and leptonic channels \cite{Kato:2020pyl}. 
Similar sensitivity is expected also for other future $e^+e^-$ collider projects \cite{deBlas:2019rxi}.

\begin{figure}[!ht]\begin{center}
	\includegraphics[width=0.6\textwidth]{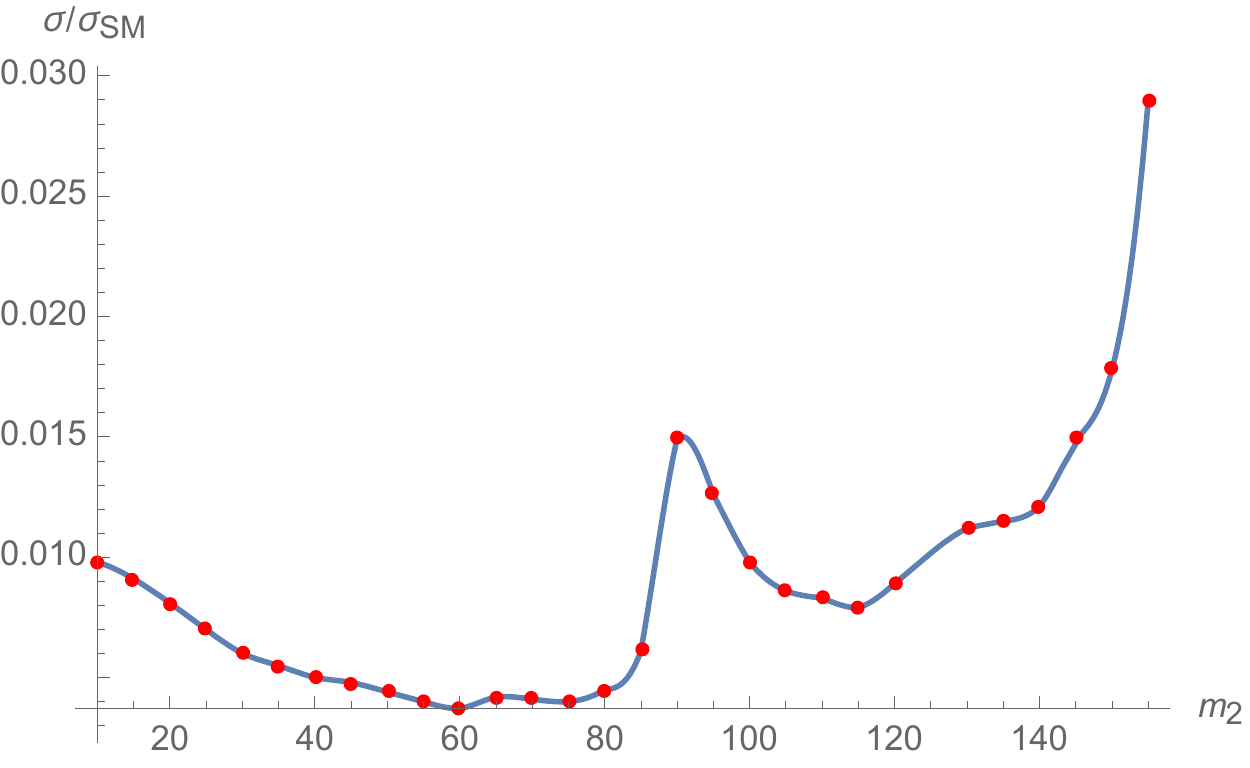}
	\caption{The 95\% CL exclusion limits \cite{Wang:2018awp,Wang:LCWS2018,Bambade:2019fyw} on the cross section for $\sigma(e^+e^-\to Z + \cdots)/\sigma_\sm$ at the ILC at $\sqrt{s}=250\gev$ as a function of the mass of the extra Higgs boson $h_2$. The ellipsis denotes an undetected final state of invariant mass $m_2$. The SM cross section assumes $m_{h_\sm}=m_2$.
Limits calculated using the CL(s) approach \cite{Read:2002hq}.}
	\label{fig:Wang_lim}
	\end{center}
\end{figure}

The Higgsstrahlung  analysis can be extended to the search for production of a generic scalar of arbitrary mass, assuming it is
produced in association with the $Z$ boson, as described in the previous section. 
The analysis procedure is the same as for the 125\,GeV Higgs boson, only the event selection criteria have to be tuned to the considered scalar mass.
The cleanest sample of Higgsstrahlung events is obtained when selecting $Z$ boson decaying into muons, as the invariant mass of the $\mu^+\mu^-$ pair can be reconstructed with sub-GeV precision and the background levels are significantly smaller than for the hadronic channel.
This channel gives the best sensitivity to the production of light scalars, below 125\,GeV, as the hadronic background levels increase rapidly towards low recoil masses, and superior recoil mass reconstruction in muon channel allows for much better suppression of non-resonant background.
No assumptions are made on the scalar decay modes or branching ratios.
The expected number of events due to SM background processes remaining after the optimized selection cuts and the corresponding
signal selection efficiency can be used to extract the expected cross-section limit on the new scalar production as a function of its mass.
Shown in figure~\ref{fig:Wang_lim} are the 95\% CL exclusion limits expected for the ILC running at $\sqrt{s}=250$\,GeV, normalized to the cross section for the SM Higgs boson production of a given mass \cite{Wang:2018awp,Wang:LCWS2018,Bambade:2019fyw}. Presented results assume $Z$-boson identification by its $\mu^+\mu^-$ decays only.
In the frequentist approach, the limit value is defined as the signal production cross section which, with probability of 95\%, would result in the observed number of events higher than the SM expectation.
The sensitivity is weakest for the scalar mass close to the mass of the $Z$ boson, due to the background from $Z$-boson pair production (with one $Z$ decaying into muons).
Scalar masses up to the kinematic limit of $\sqrt{s} - m_{Z} \sim 160$\,GeV can be probed at 250\,GeV.

To extend the limits towards higher scalar masses, $e^+ e^-$ collider running at higher energies is needed.
If the new scalar is heavier than 125\,GeV and it is expected to decay predominantly in invisible channels, the cross-section limits can be improved by considering hadronic $Z$ boson decays.
This gives increase by a factor of 20 in the expected signal event statistics (compared to the $Z\to \mu^+\mu^-$ decay channel) with only moderate increase in background levels, as the mono-$Z$ signature allows for efficient suppression of SM background processes~\cite{Mekala:2020agm}.

\section{Numerical results}
\label{num_res} 

Here we will apply the strategy described in earlier sections to investigate how large the total cross section for $Z$ 
and DM production at an $e^+e^-$ collider could be. In order to maximize the cross section we will focus on colliders running at the CoM energy close to $\sqrt{s}=250\gev$, while drawing figures we specialize to the case of
the ILC at exactly $\sqrt{s}=250\gev$~\cite{Wang:2018awp,Bambade:2019fyw}.

\begin{figure}[!ht]\begin{center}
	\includegraphics[width=0.42\textwidth]{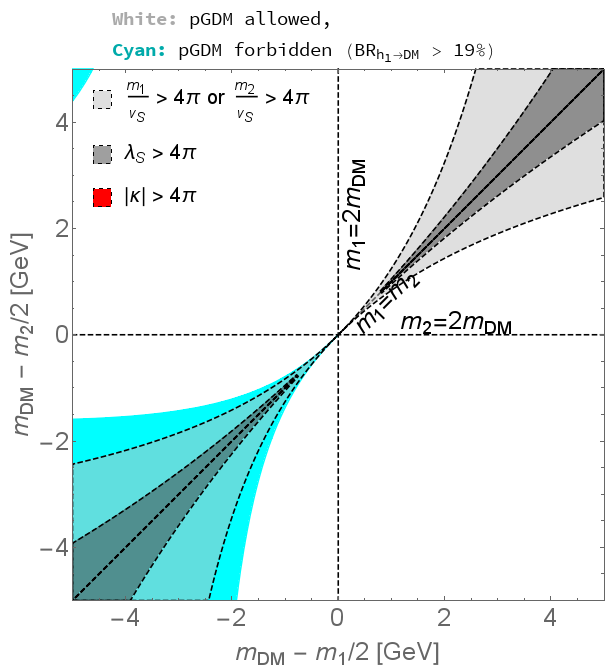}\hspace{0.1\textwidth}
	\includegraphics[width=0.42\textwidth]{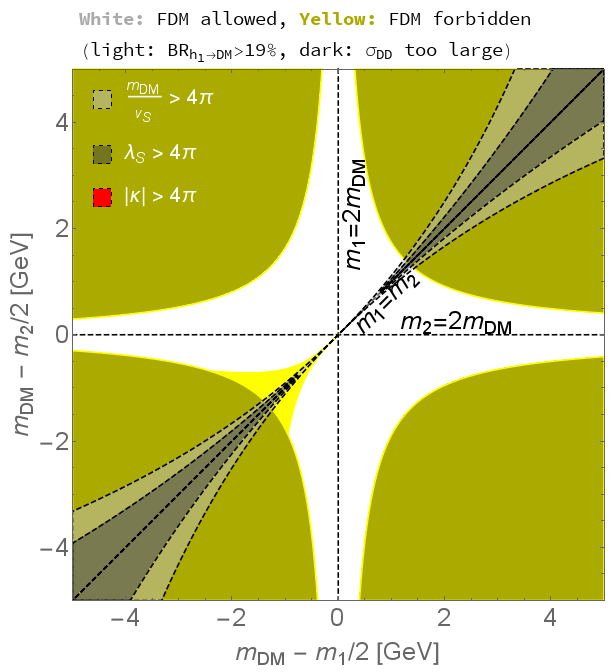}\\\vspace{4mm}
	\includegraphics[width=0.42\textwidth]{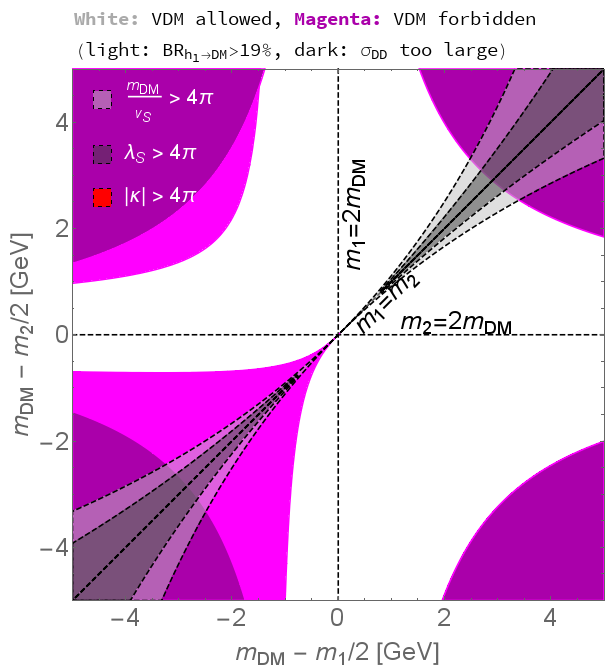}\hspace{0.1\textwidth}
	\includegraphics[width=0.42\textwidth]{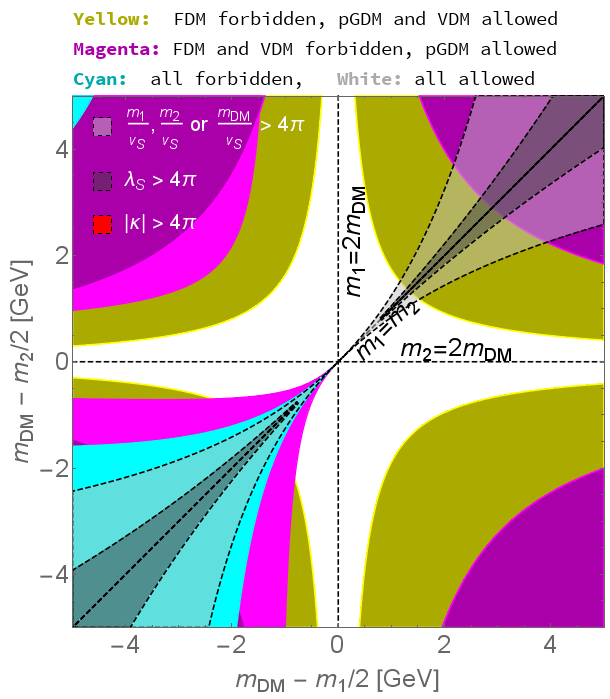}
	\caption{The parameter space allowed or forbidden for the discussed models. Top-left: pGDM model, top-right: FDM model, bottom-left: VDM model, bottom-right: the three models combined. Light- and dark-gray regions denote violation of perturbativity conditions (\ref{eq:pertCond}). Note that the $|\kappa|<4\pi$ condition is not violated in any place of the considered range of parameters.}
	\label{fig:cons_tot}
	\end{center}
	\end{figure}

\begin{figure}[!ht]\begin{center}
	\includegraphics[width=0.42\textwidth]{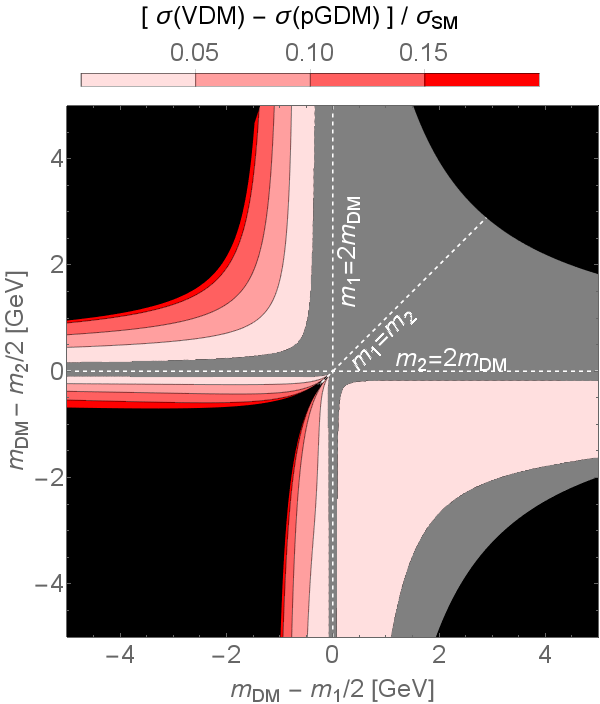}
	\caption{The difference between predictions for the pGDM and the VDM. The gray region denotes parameter space for which the difference is smaller than the limit of figure~\ref{fig:Wang_lim}. The models are compared in the region where both of them are consistent with the data, see figure~\ref{fig:cons_tot}.}
	\label{fig:SV-dif}
\end{center}\end{figure}

\begin{figure}[!ht]\begin{center}
	\includegraphics[width=0.42\textwidth]{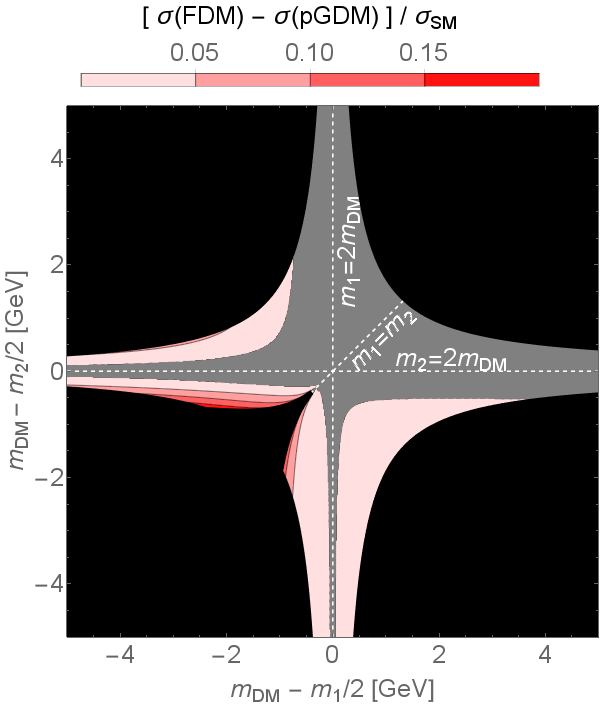}
	\hspace{0.05\textwidth}
	\includegraphics[width=0.42\textwidth]{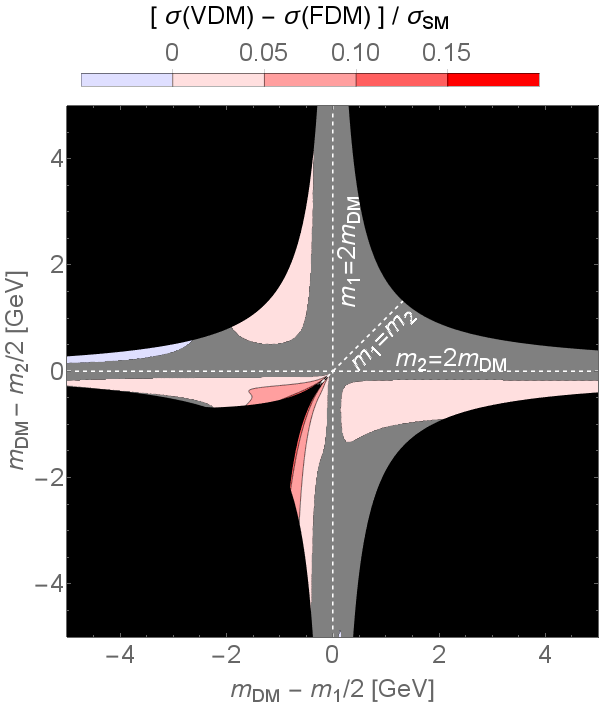}
	\caption{The difference between predictions for the pGDM and the FDM (left panel), and the FDM and the VDM (right panel). The gray region denotes parameter space for which the difference is smaller than the limit of figure~\ref{fig:Wang_lim}. The models are compared in the region where both of them are consistent with the data, 
	see figure~\ref{fig:cons_tot}.}
	\label{fig:SF-FV-dif}
	\end{center}
\end{figure}

The cross section depends on four independent variables: $m_2$, $m_\dm$, $\sin\alpha$ and $\vs$. Instead of $\vs$ one can use $\X$ defined by (\ref{X_def}), which is fixed by the relic abundance. Then, for each point $(m_2,m_\dm)$ in our plots, figures~\ref{fig:allowedSDM}--\ref{fig:SF-FV-dif}, we choose such a value of $\sin\alpha\le 0.3$ that maximizes the cross section.
Due to the resonant enhancement the cross section is much greater in the area where at least one of on-shell mediators ($h_{1,2}$) can decay into a pair of DM particles.  
As seen from (\ref{d_sig}), the differential cross section is maximized when the two Higgs bosons are on-shell at the same missing invariant mass $\sqrt{Q^2} \simeq m_1 \simeq m_2$. Therefore, the total cross section is largest when $m_1 \simeq m_2$.
Hence, the maximum appears in the lower-left quarter as close to the diagonal $m_1=m_2$ as allowed by the invisible-branching-ratio condition.\footnote{It should be remembered that in the closest vicinity of $m_1=m_2$ one should adopt the results discussed at the end of section~\ref{Production}. However, with the resolution adopted to draw plots in this paper those effects are invisible.} In the case of vector and fermion DM models, the direct detection limits on the DM-nucleon cross section are very strong, a consequence of that is that couplings between DM and the SM (parametrized by $\X$) must be severely suppressed. 
Therefore, in general, in order to satisfy the DD constraint and at the same time provide appropriate DM abundance, the early-Universe DM annihilation must occur in a vicinity of a resonance, i.e. either $2 \mdm-m_1\simeq 0$ or $2 \mdm-m_2\simeq 0$. 
For the pGDM, because of the natural suppression of the DD cross section (which is vanishing at the tree level in the limit of zero momentum transfer, see section~\ref{DD_dd_vdm}), the resonant annihilation is not necessary to reproduce the correct DM abundance. Nevertheless, to compare the models, we have found it convenient to plot the cross section in the space spanned by $\mdm-m_1/2$ and $\mdm-m_2/2$ in the vicinity of the resonance, i.e. $\mdm\simeq m_1/2$ and/or $\mdm\simeq m_2/2$. The DM and $h_2$ masses adopted hereafter satisfy the following constraints 
\beq
\left|\mdm-\frac{m_{1,2}}{2}\right|< 5\gev\,
\label{mass_range}
\eeq
what implies that $57.5\gev < m_\dm < 67.5\gev$ and $105\gev < m_2 < 145\gev$.\footnote{Region in $(m_\dm,m_2)$ plane that  corresponds to (\ref{mass_range}) is \emph{not} a rectangle.}

In figures~\ref{fig:allowedSDM}--\ref{fig:allowedFDM} we plot maximized cross section for the $Z$ and DM production (normalized to the SM prediction for the $Zh_\sm$ production, i.e. $\sigma_\sm=\sigma(e^+e^-\to Z h_\sm)|_{m_{h_\sm}=m_1}$) at the ILC for pGDM, VDM and FDM models, respectively. The greenish colors denote regions where the models satisfy adopted constraints showing (by color) the corresponding cross section. The cyan marks regions excluded by the SM invisible BR limit, $\br(h_1\to\dm)<0.19$, while the black corresponds to parameters excluded by the DD limit (see section~\ref{DD_dd_vdm}). As explained earlier, the allowed regions for the VDM and the FDM models appear in the vicinity of resonant DM annihilation. For the gray region, the expected $95\%$ CL sensitivity limit for $\sigma/\sigma_\sm$ (shown in figure~\ref{fig:Wang_lim}) is above the $\sigma/\sigma_\sm$ prediction. Therefore, one can conclude that the greenish regions are those which are detectable at the ILC, assuming that the $\sin\alpha$ is close to the value that maximizes the cross section. It turns out that usually the $\sin\alpha$ that provides maximal cross section is just at the largest value allowed by the LHC Higgs signal measurement, $\sin\alpha \simeq 0.3$. The fair conclusion from inspecting figures~\ref{fig:allowedSDM}--\ref{fig:allowedFDM} is that in the substantial part of the parameter range that was shown, the DM production can be detected at future $e^+e^-$ colliders running around $\sqrt{s}=250\gev$.

\begin{minipage}{\textwidth}
\begin{figure}[H]\begin{center}
	\hspace*{-2cm}
	\raisebox{-0.5\height}{
		\includegraphics[width=0.42\textwidth]{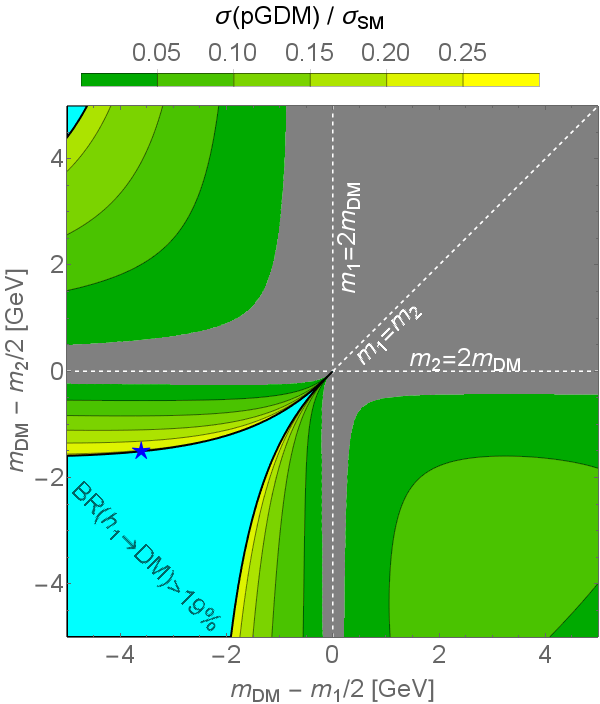}
		}
	\hspace{0.5cm}
	\raisebox{-0.5\height}{
	\scalebox{0.8}{\begin{minipage}{0.4\textwidth}\begin{gather*}
	\text{\underline{benchmark point for pGDM}}\\
	\begin{aligned}
	&m_2=120.8\gev\;,&&m_\dm=58.9\gev\;,\\
	&\sin\alpha=0.30\;,&&\vs=646\gev\;,\\
	&\Gamma_1=7.4\cdot10^{-3}\gev\;,&&\Gamma_2=9.8\cdot10^{-3}\gev\;,\\
	&\br(h_1\to\dm)=19\%\;,&&\br(h_2\to\dm)=95\%\;,
	\end{aligned}\\
	\sigma=62\text{ fb}
	\end{gather*}\end{minipage}}}
	\caption{The figure shows, for the pGDM, the allowed region (greenish), the region forbidden by the invisible BR of $h_1$ (cyan) where $\br(h_1\to \dm)>19\%$ and the gray region where the normalized cross section falls below its expected precision at the $95\%$ CL shown in figure~\ref{fig:Wang_lim}. Coloring of the greenish area, explained in the legend, shows the value of the normalized total cross section $\sigma/\sigma_{SM}$. The star denotes the chosen benchmark point, characterized by relatively high cross section.}
	\label{fig:allowedSDM}
	\end{center}
\end{figure}

\begin{figure}[H]\begin{center}
\begin{minipage}{0.42\textwidth}\begin{center}
	\includegraphics[width=\textwidth]{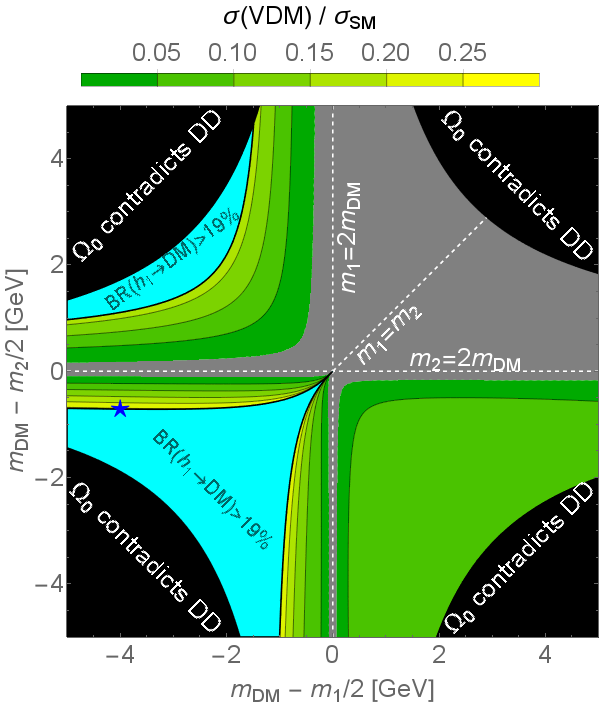}
	\scalebox{0.8}{\begin{minipage}{\textwidth}\begin{gather*}
	\text{\underline{benchmark point for VDM}}\\
	\begin{aligned}
	&m_2=118.4\gev\;,&&m_\dm=58.5\gev\;,\\
	&\sin\alpha=0.30\;,&&\vs=561\gev\;,\\
	&\Gamma_1=7.4\cdot10^{-3}\gev\;,&&\Gamma_2=6.4\cdot10^{-3}\gev\;,\\
	&\br(h_1\to\dm)=18\%\;,&&\br(h_2\to\dm)=92\%\;,
	\end{aligned}\\
	\sigma=61\text{ fb}
	\end{gather*}\end{minipage}}
\end{center}\end{minipage}
\begin{minipage}{0.1\textwidth}~\end{minipage}
\begin{minipage}{0.42\textwidth}\begin{center}
	\includegraphics[width=\textwidth]{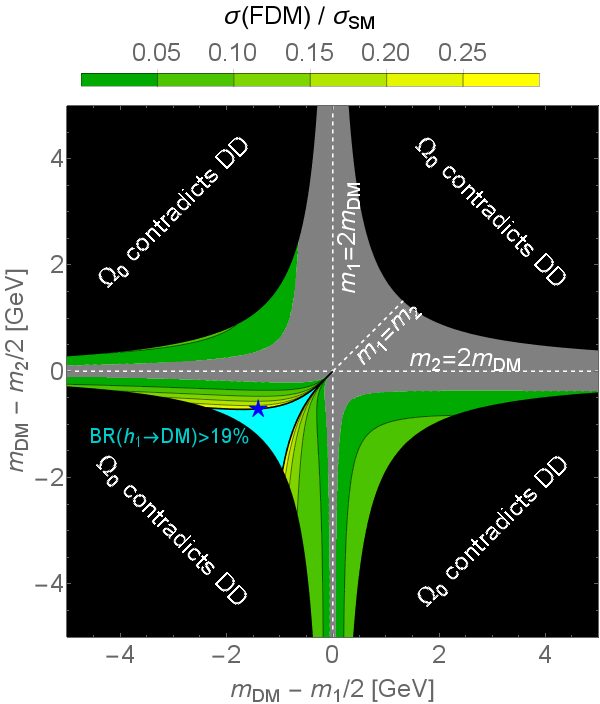}
	\scalebox{0.8}{\begin{minipage}{\textwidth}\begin{gather*}
	\text{\underline{benchmark point for FDM}}\\
	\begin{aligned}
	&m_2=123.6\gev\;,&&m_\dm=61.1\gev\;,\\
	&\sin\alpha=0.30\;,&&\vs=76\gev\;,\\
	&\Gamma_1=7.4\cdot10^{-3}\gev\;,&&\Gamma_2=5.9\cdot10^{-3}\gev\;,\\
	&\br(h_1\to\dm)=18\%\;,&&\br(h_2\to\dm)=91\%\;,
	\end{aligned}\\
	\sigma=59\text{ fb}
	\end{gather*}\end{minipage}}
	\end{center}\end{minipage}
	\caption{As in figure~\ref{fig:allowedSDM} for the VDM model (left) and the FDM model (right). The region forbidden by the DD constraint is denoted by black color.}
	\label{fig:allowedVDM}
	\label{fig:allowedFDM}
\end{center}\end{figure}
\end{minipage}
  
The simplest and straightforward way to disentangle the models is to measure $m_2$ and $\mdm$ and then verify if the measured masses are consistent with any of the discussed models after imposing constraints. In other words, one would need to check if for the measured values of $m_2$ and $\mdm$ the corresponding point $(\mdm-m_1/2, \mdm-m_2/2)$  is located in the greenish area in any of figures~\ref{fig:allowedSDM}--\ref{fig:allowedFDM}. In order to facilitate and illustrate the verification, we have made  plots shown in figure~\ref{fig:cons_tot}.
The upper-left panel shows the cyan region where the pGDM is excluded by the $h_1$-invisible-BR condition, while the white region is allowed. In the upper-right panel the yellow regions denote region where the FDM model is disallowed, while white color stands again for region that agrees with all the constraints. Similarly, the lower-left panel shows forbidden (magenta) and allowed (white) regions for the VDM model. The lower-right plot combines results for all the models; again, white denotes the region where all the models are allowed. As it is seen, there exist regions where two or even three models coexist. However, there is also, in the lower-right panel, the magenta region where only the pGDM may exist. Therefore, the very first step in an attempt to disentangle the models should be a measurement of $m_2$ and $\mdm$ and its verification against the results shown in figure~\ref{fig:cons_tot}. Even though there is a substantial region of full degeneracy (white), there exists also significant area where some valuable conclusions could be drawn. It is even conceivable that this measurement would be consistent with the spin 0 (pGDM) hypothesis only.

Now, we would like to focus on estimating chances to disentangle pairs of the models by the measurement of the normalized cross section $\sigma/\sigma_\sm$. In order to verify this option, we plot (figures~\ref{fig:SV-dif} and \ref{fig:SF-FV-dif}) differences between model predictions and compare them against the expected experimental precision given by the limit provided by figure~\ref{fig:Wang_lim}. As previously, it turns out that the highest differences are obtained for $\sin\alpha$ as large as allowed, i.e. $\sin\alpha\simeq0.3$. More reddish color indicate parameter regions for which models that are being compared are easier to disentangle since there an absolute value of the corresponding difference of cross sections is larger. It is clear that the disentanglement is a very ambitious task. It seems that only the VDM and the pGDM could be relatively easily disentangled by the measurement of $e^+e^-\to Z+\cdots$ cross section at future $e^+e^-$ colliders operating near $\sqrt{s}=250\gev$ if the parameters are in the more reddish regions of figure~\ref{fig:SV-dif}. 


\section{Summary}
\label{sec:summary}
In this analysis, our goal was to investigate how could one disentangle models of dark matter of different spin at future $e^+e^-$ colliders operating near $\sqrt{s}=250\gev$. For that purpose, we adopted the ILC project with the CoM energy at $\sqrt{s}=250\gev$. 
Our strategy was pragmatic and phenomenological. We considered three nearly simplest models of dark matter of spin $0$, $1$ and $\half$. The models adopted here were not the "simplified" ones, discussed often in a phenomenological literature on dark matter; in contrast, they were simple but attractive, consistent and renormalizable quantum field theories. In spite of dark-matter-spin differences, the models considered here share exactly the same parameter space, so the comparison point by point was meaningful. It turned out that the most promising region of the parameter space is located near the double resonance $2 m_\dm \sim m_{1,2}$. It has been shown that in this region, in the closest vicinity of $m_1=m_2$, the naive Breit-Wigner strategy must be replaced by a proper resummation of 1-loop Higgs-boson self-energies that takes into account their off-diagonal elements.

In order to verify if a model was testable, we had adopted expected 95\% CL sensitivity for the measurement of the $e^+e^-\to Z + \cdots$ cross section obtained for the ILC project. It has been assumed that only the $Z$ boson is reconstructed without any other detector activity. Predictions of the models were calculated taking into account the dark matter abundance, indirect and direct detection experiments and the collider constraints on the Higgs-boson invisible branching ratio and limits on the mixing angle (present in all the models). That way, regions of the parameter space where the cross section would be measurable were obtained for each of the models. We have also discussed the possibility to disentangle the models by a measurement of the cross section.  It turned out that the most optimistic case is the detection of differences between the pseudo-Goldstone dark matter (spin 0) and the vector dark matter (spin 1). Regions of the parameter space where no model is allowed or some models could coexist were also determined.     

\noi {\bf Acknowledgments}\\

\noi The work was partially supported by the National Science Centre (Poland) OPUS research projects under contracts nos. UMO-2017/25/B/ST2/00191 and UMO-2017/25/B/ST2/00496, and HARMONIA project under contract no. UMO-2015/18/M/ST2/00518 (2016--2019).

\appendix
\section{Higgs boson self-energies and decay widths}\label{propagator}
Here we collect results for imaginary parts of two-point functions $\Pi_{ij}(Q^2)$ for Higgs bosons 
$h_{i,j}$. Each self-energy is a sum of contributions of loops with 
various intermediate states\footnote{We omit tadpole and seagull diagrams.} being on-shell:
	\beq
	\Pi_{ij}=\Pi_{ij}^{\dm}+\Pi_{ij}^{W^+W^-}\!\!\!+\Pi_{ij}^{ZZ}
	+\sum_{q}\Pi_{ij}^{q\bar q}+\sum_{l}\Pi_{ij}^{l^+l^-}+\sum_{k,l}\Pi_{ij}^{h_kh_l}\;,
	\eeq
where $\dm$ stands for dark matter particle $A$, $X$ or $\psi$ while $q$ denotes SM quarks and $l$ denotes SM leptons. These contributions are given by (see \cite{Duch:2018ucs} for the VDM case\footnote{In \cite{Duch:2018ucs}, there is an additional $i$ factor in the definition of $V_{ijk}$, hence additional minus in their version of eq.~(\ref{eq:Pihh}).}):
	\bal
	\Pi^{\dm}_{ij}(Q^2)&=I(Q^2,m_\dm,m_\dm)\frac{\R_{2i}\R_{2j}}{32\pi^2\vs^2}(m_im_j)^2\times\notag\\
	&\quad\times\begin{cases}
	1	&(\sdm)\\
	1-2\,m_\dm^2\frac{4Q^2-m^2_i-m^2_j}{(m_im_j)^2}+12\left(\frac{m_\dm^2}{(m_im_j)^2}\right)^2	&(\vdm)\\
	2\frac{m_\dm^2Q^2}{(m_im_j)^2}\left(1-4\frac{m_\dm^2}{Q^2}\right)	&(\fdm)
	\end{cases}\;,\\
	\Pi^{W^+W^-}_{ij}(Q^2)&=I(Q^2,m_W,m_W)\frac{\R_{1i}\R_{1j}}{16\pi^2v^2}(m_im_j)^2\times\notag\\
	&\quad\times\left[1-2m_W^2\frac{4Q^2-m^2_i-m^2_j}{(m_im_j)^2}+12\frac{m_W^4}{(m_im_j)^2}\right]\;,\\
	\Pi^{ZZ}_{ij}(Q^2)&=I(Q^2,m_Z,m_Z)\frac{\R_{1i}\R_{1j}}{32\pi^2v^2}(m_im_j)^2\times\notag\\
	&\quad\times\left[1-2m_Z^2\frac{4Q^2-m^2_i-m^2_j}{(m_im_j)^2}+12\frac{m_Z^4}{(m_im_j)^2}\right]\;,\\
	\Pi^{q\bar q}_{ij}(Q^2)&=I(Q^2,m_q,m_q)\cdot
	\frac{3\R_{1i}\R_{1j}}{8\pi^2v^2}m_q^2Q^2\left(1-4\frac{m_q^2}{Q^2}\right)\;,\\
	\Pi^{l^+l^-}_{ij}(Q^2)&=I(Q^2,m_l,m_l)\cdot
	\frac{\R_{1i}\R_{1j}}{8\pi^2v^2}m_l^2Q^2\left(1-4\frac{m_l^2}{Q^2}\right)\;,\\
	\Pi^{h_k h_l}_{ij}(Q^2)&=I(Q^2,m_k,m_l)\cdot
	\frac{V_{ikl}V_{jkl}}{32\pi^2}\;,\label{eq:Pihh}
	\eal
where
	\bal
	V_{111}&\equiv 3m_1^2\left(\frac{\sin^3\alpha}{\vs}+\frac{\cos^3\alpha}{v}\right)\;,\\
	V_{112}=V_{121}=V_{211}&\equiv
		(2m_1^2+m_2^2)\sin\alpha\cos\alpha\left(\frac{\sin\alpha}{\vs}-\frac{\cos\alpha}{v}\right)\;,\\
	V_{221}=V_{212}=V_{122}&\equiv
		(m_1^2+2m_2^2)\sin\alpha\cos\alpha\left(\frac{\cos\alpha}{\vs}+\frac{\sin\alpha}{v}\right)\;,\\
	V_{222}&\equiv 3m_2^2\left(\frac{\cos^3\alpha}{\vs}-\frac{\sin^3\alpha}{v}\right)
	\eal
are the couplings corresponding to the $h_ih_jh_k$ vertices ($i,j,k=1,2$)~\cite{Duch:2018ucs}
and
	\beq\begin{aligned}
	I(Q^2,m_a,m_b)&\equiv i\cdot\Im\left[B_0(Q^2,m_a^2,m_b^2)\right]=\\
	&=i\cdot\Im\left[\frac{1}{i\pi^2}\int\frac{d^4l}{(l^2-m^2_a)[(l+Q)^2-m^2_b]}\right]=\\
	&=i\pi\cdot\frac{\lambda^{1/2}(Q^2,m_a^2,m_b^2)}{Q^2}\cdot\one{Q^2>(m_a+m_b)^2}\\
	\Rightarrow\quad I(Q^2,m,m)&=i\pi\cdot\sqrt{1-\frac{4m^2}{Q^2}}\cdot\one{Q^2>4m^2}
	\end{aligned}\eeq
are imaginary parts\footnote{The choice of the sign depends on the corresponding choice in $\ln(-1)=\pm i\pi$. We want the imaginary part to be positive, since it corresponds to the correct asymptotic value, i.e. $\Pi_{ii}(m_i^2)=+im_i\Gamma_i$.} (times $i$) of appropriate loop integrals $B_0$ \cite{Denner:1991kt}, where $\lambda$ denotes the K\"all\'en function, defined as
	\beq
	\lambda(a,b,c)\equiv a^2+b^2+c^2-2(ab+bc+ca)\;.
	\eeq
By straightforward calculations it can be shown that
	\beq\begin{aligned}
	\left[\left(\Pi^{ab}_{11}-\Pi^{ab}_{22}\right)
	-\left(\tan\alpha\cdot\Pi^{ab}_{12}-\cot\alpha\cdot\Pi^{ab}_{21}\right)
	\right]\at{m_1=m_2}=0
	\end{aligned}\eeq
for $ab=\dm,\,W^+W^-,\,ZZ,\,q\bar q,\,l^+l^-,\,h_kh_l$ (in the last case one has to sum over $k,l=1,2$). Hence, also the sum over all contributions vanishes in this limit:
	\beq\begin{aligned}
	\left[\left(\Pi_{11}-\Pi_{22}\right)
	-\left(\tan\alpha\cdot\Pi_{12}-\cot\alpha\cdot\Pi_{21}\right)
	\right]\at{m_1=m_2}=0\;.
	\end{aligned}\eeq

The $h_1$'s and $h_2$'s partial widths can be calculated as
	$$\Gamma_{h_i\to ab}=\frac{\Pi_{ii}^{ab}(m_i^2)}{im_i}\;.$$
The widths relevant for this project are therefore given by 
	\bal
	\Gamma_{h_i\to\dm}&=\frac{\R_{2i}^2}{\vs^2}
	\frac{m_i^3}{32\pi}\sqrt{1-\frac{4m_\dm^2}{m_i^2}}\times\notag\\
		&\quad\times\begin{cases}
		1 & (\sdm) \\
		1 - 4\frac{m_\dm^2}{m_i^2} + 12 \left(\frac{m_\dm^2}{m_i^2}\right)^2 & (\vdm)\\
		2\frac{m_\dm^2}{m_i^2}\left(1-4\frac{m_\dm^2}{m_i^2}\right) & (\fdm)
		\end{cases}\;,\\
		\Gamma_{h_i\to\sm}&=\R_{1i}^2\cdot\gamma(m_i)\\
		&\text{($\gamma$ denotes the decay width of SM Higgs particle of given mass)}\;,\notag\\
		\Gamma_{h_1\to h_2h_2}&=\sin^2\alpha\,\cos^2\alpha\;(m_1^2+2m_2^2)^2
		\left(\frac{\cos\alpha}{\vs}+\frac{\sin\alpha}{v}\right)^2
		\frac{\sqrt{m_1^2-4m_2^2}}{32\pi m_1^2}\simeq\notag\\
		&\simeq\;\frac{\sin^2\alpha\,\cos^4\alpha}{\vs^2}
		(m_1^2+2m_2^2)^2\frac{\sqrt{m_1^2-4m_2^2}}{32\pi m_1^2}\;,\\
		\Gamma_{h_2\to h_1h_1}&=\sin^2\alpha\,\cos^2\alpha\;(2m_1^2+m_2^2)^2
		\left(\frac{\sin\alpha}{\vs}-\frac{\cos\alpha}{v}\right)^2
		\frac{\sqrt{m_2^2-4m_1^2}}{32\pi m_2^2}\simeq\notag\\
		&\simeq\;\frac{\sin^2\alpha\,\cos^4\alpha}{v^2}
		(2m_1^2+m_2^2)^2\frac{\sqrt{m_2^2-4m_1^2}}{32\pi m_2^2}\;.
	\eal
\section{Passarino-Veltman functions}\label{sec:pave}
Functions used in eq.~(\ref{loop_res}) are defined in terms of Passarino-Veltman functions \cite{Denner:1991kt}:
	\bal
	D(0,0,\sqrt{p^2};m_a,m_b,m_c,m_d)&\equiv\frac{p^\mu}{p^2}D_\mu(0,0,p;m_a,m_b,m_c,m_d)\;,\\
	C(0,\sqrt{p^2};m_a,m_b,m_c)&\equiv\frac{p^\mu}{p^2}C_\mu(0,p;m_a,m_b,m_c)\;.
	\eal
Explicit values are:
	\bal
	&D(0,0,m_\dm;m_1,m_1,m_2,m_\dm)=\\
	&\qquad=\frac{p^\mu}{m_\dm^2}\frac{1}{i\pi^2}\int d^4l
		\frac{l_\mu}{(l^2-m_1^2)^2(l^2-m_2^2)\left[(l+p)^2-m_\dm^2\right]}\at{p^2=m_\dm^2}=\notag\\
	&\qquad=\frac{1}{m_1^2-m_2^2}\left[
			 C(0,m_\dm;m_1,m_1,m_\dm)
			-C(0,m_\dm;m_1,m_2,m_\dm)
		\right]\;,\notag\\
	&D(0,0,m_\dm;m_1,m_2,m_2,m_\dm)=\\
	&\qquad=\frac{p^\mu}{m_\dm^2}\frac{1}{i\pi^2}\int d^4l
		\frac{l_\mu}{(l^2-m_1^2)(l^2-m_2^2)^2\left[(l+p)^2-m_\dm^2\right]}\at{p^2=m_\dm^2}=\notag\\
	&\qquad=-\frac{1}{m_1^2-m_2^2}\left[
			 C(0,m_\dm;m_2,m_2,m_\dm)
			-C(0,m_\dm;m_1,m_2,m_\dm)
		\right]\;,\notag\\
	&\label{eq:Cmu}C(0,m_\dm;m_1,m_2,m_\dm)=\\
	&\qquad=\frac{p^\mu}{m_\dm^2}\frac{1}{i\pi^2}\int d^4l
		\frac{l_\mu}{(l^2-m_1^2)(l^2-m_2^2)\left[(l+p)^2-m_\dm^2\right]}\at{p^2=m_\dm^2}=\notag\\
	&\qquad=\frac{1}{m_1^2-m_2^2}\left[
			 B(m_\dm;m_1,m_\dm)
			-B(m_\dm;m_2,m_\dm)
		\right]\;,\notag
	\eal
where the following auxiliary functions are used:
	\bal
	&C(0,m_\dm;m_i,m_i,m_\dm)=\\
	&\qquad=-\frac{1}{m_\dm^2}
		\left[1
		+\frac{x_i^+(x_i^+-1)}{x_i^+-x_i^-}\ln\left(\frac{x_i^+-1}{x_i^+}\right)
		-\frac{x_i^-(x_i^--1)}{x_i^+-x_i^-}\ln\left(\frac{x_i^--1}{x_i^-}\right)\right]\;,\notag\\
	&\label{eq:Bmu}B(m_\dm;m_i,m_\dm)=\\
	&\qquad=-\frac{1}{2}
		\left[\left(\frac2\epsilon-\gamma+\ln\frac{\mu^2}{m_\dm^2}\right)+\right.\notag\\
		&\mkern100mu\left.+1+\frac{m_i^2}{m_\dm^2}
		+(x_i^+)^2\ln\left(\frac{x_i^+-1}{x_i^+}\right)
		+(x_i^-)^2\ln\left(\frac{x_i^--1}{x_i^-}\right)\right]\;,\notag\\
	&\bald
	x_i^\pm&\equiv\frac{m_i^2\pm\sqrt{m_i^4-4m_i^2m_\dm^2}}{2m_\dm^2}\;.
	\eald
	\eal
The $\left(\frac2\epsilon-\gamma+\ln\frac{\mu^2}{m_\dm^2}\right)$ term present in eq.~(\ref{eq:Bmu}) appears due to the chosen regularization  scheme and cancels out in eq.~(\ref{eq:Cmu}).


\bibliography{bib_ep_em_DM}{}
\bibliographystyle{JHEP}

\end{document}